\documentclass[11pt,onecolumn]{article}

\usepackage{graphicx}% Include figure files
\usepackage{dcolumn}% Align table columns on decimal point
\usepackage{bm}% bold math
\usepackage{enumitem}

\usepackage[
    a4paper,
    left=2.5cm,
    right=2.5cm,
    top=1.5cm,
    bottom=1.5cm
]{geometry}
\usepackage{fancyhdr}
\usepackage{enumitem}
\usepackage{amssymb}

\usepackage[compact]{titlesec}
\titleformat{\section}
  {\sffamily\large\bfseries}
  {\thesection}{0.6em}{}
  
  \titleformat{\subsection}
  {\sffamily\normalsize\bfseries}
  {\thesubsection}{0.6em}{}

\usepackage[T1]{fontenc}
\usepackage{lmodern}

\usepackage{lineno} 

\usepackage{setspace}
\titleformat{\section}
  {\large\bfseries}
  {}{0pt}{}

\titleformat{\subsection}
  {\normalsize\bfseries}
  {}{0pt}{}

\titlespacing*{\section}{0pt}{16pt}{6pt}
\titlespacing*{\subsection}{0pt}{12pt}{4pt}

\usepackage[
    format=plain,
    justification=justified,
    singlelinecheck=false,
    font={small,sf},
    labelfont=bf,
    labelsep=space
]{caption}

\usepackage{graphicx}
\usepackage{booktabs}
\usepackage{tabularx}
\usepackage{float}

\usepackage{amsmath}

\usepackage[super,sort&compress,comma]{natbib}
\usepackage[version=3]{mhchem}

\usepackage{xcolor}
\usepackage{hyperref}

\newcommand{\rb}{\mathbf r}
\newcommand{\xb}{\mathbf x}
\newcommand{\pb}{\mathbf p}
\newcommand{\ub}{\mathbf u}

\newcommand{\Bb}{\mathbf B}
\newcommand{\Fb}{\mathbf F}
\newcommand{\Ub}{\mathbf U}
\newcommand{\Tb}{\mathbf T}

\newcommand{\ehat}{\hat{\mathbf{e}}}
\newcommand{\Bhat}{\hat{\mathbf{B}}}
\newcommand{\phat}{\hat{\mathbf{p}}}
\newcommand{\rhat}{\hat{\mathbf r}}

\newcommand{\nablab}{\boldsymbol{\nabla}}
\newcommand{\xib}{\boldsymbol{\xi}}

\begin{document}
\begin{tabular}{m{0.2cm} p{16.5cm} }
\vspace{0.3cm} & \vspace{0.3cm}\\
 \vspace{0.3cm} & \vspace{0.3cm}\\
 
 & \noindent\LARGE{\textbf{Self-Organized Antipodal Multifunctional Micromotors via Evaporative Liquid–Liquid Phase Separation    }} \\%Article title goes here instead of the text "This is the title"
\vspace{0.3cm} & \vspace{0.3cm} \\

 & \noindent\large{Senthan Pugalneelam Parameswaran$^\dag$$^1$, Akshay Sidhi$^\dag$$^1$,  Ambareesh Shrivastav$^\dag$$^1$, Dibyendu Das$^2$, Tapan Chandra Adhyapak$^*$$^1$, Dileep Mampallil$^*$$^1$ } \\
 
 & \vspace{0.3cm} \\
 
&  1. \textit{Indian Institute of Science Education and Research Tirupati, Yerpedu AP, 517619 INDIA.; E-mail: dileep.mampallil@iisertirupati.ac.in, adhyapak@iisertirupati.ac.in} \\
&  \\

& 2. \textit{Indian Institute of Science Education and Research Kolkata, Mohanpur, Haringhata Farm, West Bengal 741246 INDIA.} \\
 
 & \noindent\normalsize{

Active microparticles with spatially organized multifunctionality could enable new forms of autonomous transport and chemical activity, yet creating such architectures within individual particles remains challenging. Here, we show that such complex functional architectures emerge spontaneously when multiple functional components (enzymes, nanoparticles, etc.) are introduced into evaporating multicomponent droplets undergoing liquid–liquid phase separation. Coupled Marangoni flows and interfacial dynamics drive the asymmetric localization of functional nanoparticles, positioning distinct functionalities on opposite sides of a microparticle in an architecture we term \textit{antipodal}. This organization generates emergent magnetic anisotropy when magnetic nanoparticles are incorporated, enabling catalytically propelled motion parallel or antiparallel to a uniform magnetic field without an external chemical gradient. We also demonstrate enzyme–enzyme antipodal micromotors that combine weak chemotactic steering with strong propulsion driven by a spatially uniform chemical fuel, a strategy relevant to environments where chemical gradients are weak. Extensive theoretical modelling elucidates the underlying interplay of Marangoni flows and interfacial dynamics, providing a mechanistic basis for tunability. Our findings establish a route to spatially programmed active matter, in which functional architectures and directional capabilities emerge spontaneously through coupled hydrodynamic and interfacial dynamics.
 } 

\end{tabular}

%%%FONT SETUP - please do not change any commands within this section
\renewcommand*\rmdefault{bch}\normalfont\upshape
\sffamily
% \end{@twocolumnfalse} \vspace{0.6cm}
 
% ]

%%%END OF TITLE, AUTHORS AND ABSTRACT%%%

%%%FONT SETUP - please do not change any commands within this section
\renewcommand*\rmdefault{bch}\normalfont\upshape
\rmfamily
\section*{}
\vspace{-1cm}

%%%FOOTNOTES%%%

%\footnotetext{\dag~Supplementary Information available: [details of any supplementary information available should be included here]. See DOI: 10.1039/cXsm00000x/} 
\footnote{\dag~Equal contribution}

%%%END OF FOOTNOTES%%%

\newpage

\vspace*{1.0cm}
\section{Introduction}

Integrating multiple functional components into micro- and nanomotors is essential for applications ranging from autonomous targeted delivery \cite{pantarotto2008autonomous, patra2013intelligent, arque2022enzyme, li2023medical, simo2024urease, chen2025roadmap, choi2024urease} and the construction of synthetic biomimetic systems \cite{peters2014cascade, wang2018biomimetic, wang2023cell} and active droplets \cite{PRL130.128401, ZwickerNatPhy} to programmable active matter \cite{Vutukuri2026program}. In targeted delivery, for example, multifunctionality enables motors to combine therapeutic payloads with agents that facilitate tumour-tissue penetration \cite{cao2022oral, serra2024catalase}, integrate multiple enzymes \cite{dey2015micromotors, arque2019intrinsic, ghosh2023dual} to enhance chemotaxis in weak chemical gradients \cite{ ghosh2023dual}, and incorporate magnetic components for externally guided navigation \cite{tang2026enzymatic, zhang2021dual, naikwadi2024self, chen2025roadmap}. However, functionality alone is not sufficient: the spatial organization of these components within individual motors can be crucial for coupling propulsion, directional steering, and cargo delivery. Achieving such architectures remains challenging because conventional fabrication strategies \cite{kirillova2019hybrid, joye2014biopolymer, carvalho2023multifunctional} offer limited control over complex spatial arrangements and may compromise biomolecular activity. Moreover, magnetic steering often relies on engineered shape anisotropy of magnetic components, adding further fabrication complexity \cite{zhou2021magnetically}. A simple strategy that spontaneously organizes multiple functional components while establishing magnetic anisotropy could therefore enable multifunctional micromotors with enhanced directional control and autonomous transport.

\begin{center}
	\begin{figure}[htbp] 
		\centering
		\includegraphics[scale= 0.99]{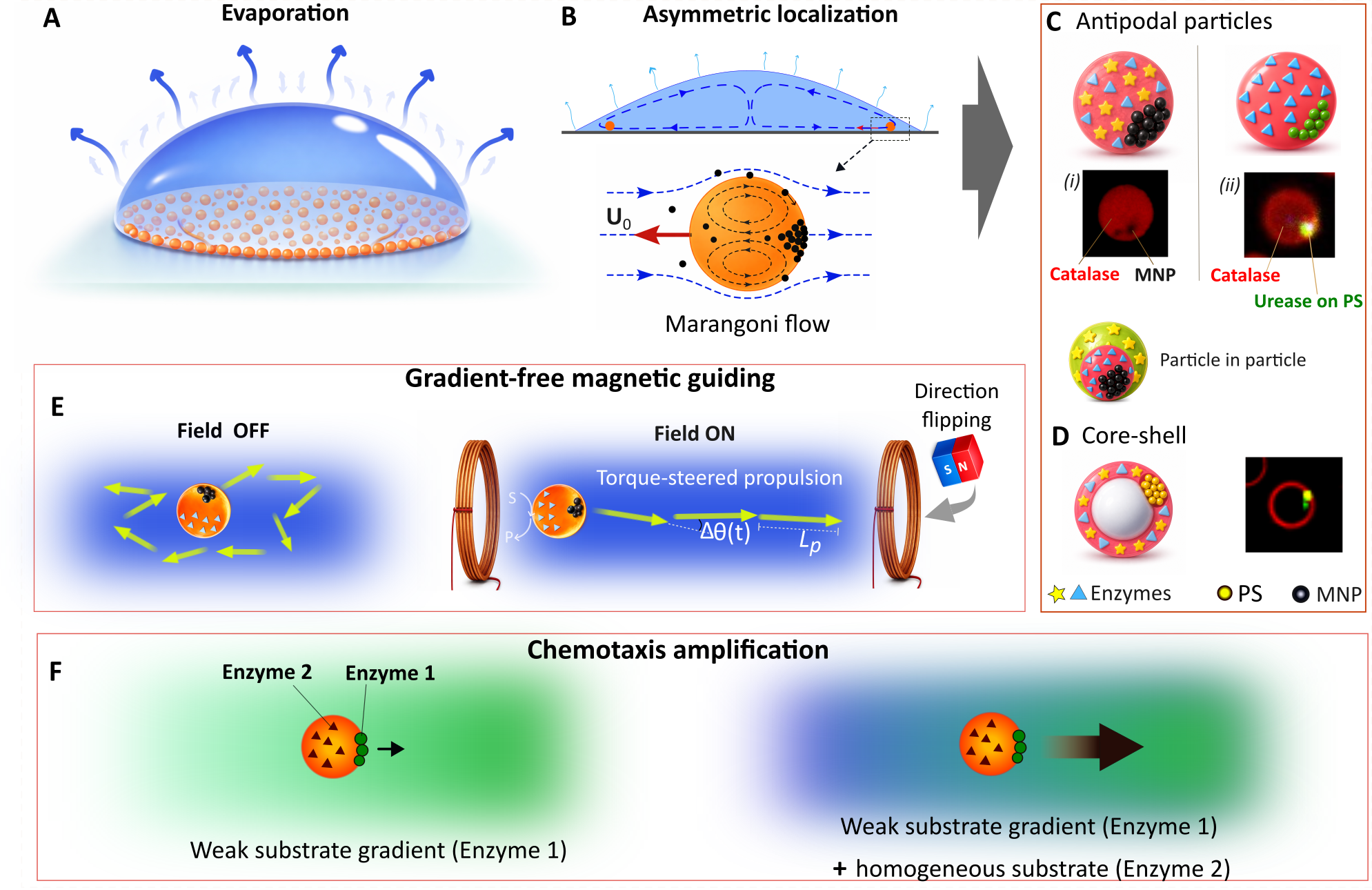}
		\caption{
\textbf{Fabrication and dual-driving mechanism of multifunctional particles}. \textbf{A} Evaporation-induced liquid–liquid phase separation forms microcondensates that encapsulate enzymes and nanoparticles. \textbf{B} Marangoni flows and interfacial capture generate asymmetric localization of nanoparticles within the condensates. \textbf{C, D} Stable microparticles, particle-in-particle architectures, and core–shell structures with antipodally (situated on opposite sides) organized magnetic and chemical functionalities are prepared.  Representative images of microparticles (diameter: 6 $\mu$m) with asymmetrically localized (i) magnetite (MNP) and (ii) urease-coated polystyrene (PS) particles (1 $\mu$m). \textbf{E} \textbf{Application 1}: The emergent magnetic anisotropy facilitates torque-induced steering, a weak net magnetic dipole moment facilitates flipping the direction by an external magnet, and multifunctionality facilitates simultaneous catalytic propulsion. \textbf{F} \textbf{Application 2}: (Left) Enzyme 1 induces chemotaxis in a weak gradient of its substrate (say, urea for urease). (Right) The presence of a strong, homogeneously present substrate of enzyme 2 propels the micromotor strongly, amplifying the chemotaxis, thanks to the antipodal positioning of the enzymes. }  \label{fig1}
	\end{figure}
\end{center}

Here, we show that liquid–liquid phase separation (LLPS) \cite{naz2024self, park2025pompoms} in evaporating multicomponent aqueous droplets \cite{wakata2025evaporation, mampallil2025evaporation, Guo2021, May2022, watanabe2022evaporation, Rai2024, pnas.1602260113, kumar2024evaporative} provides a robust, bottom-up, one-step strategy to fabricate multifunctional microcondensates that can be stabilized to form particle and core-shell \cite{ivanov2025dynamic} micromotors of protein (bovine serum albumin), polymer (chitosan), and hybrid hierarchical matrix (Fig.\ref{fig1}). During LLPS, the spontaneous and preferential encapsulation of biomolecules such as enzymes, peptides, and nucleic acids \cite{dave2025adaptive, keating2012aqueous, ramos2022protein, sakuta2020self, Guo2021, Qi2024, hu2023liquid} is exploited to induce multifunctionality simultaneously.

The functional architecture of these microcondensates emerges spontaneously during evaporation. The coupling between the Marangoni-driven flows and interfacial adsorption captures and positions nanoparticles asymmetrically in the microcondensates. It results in the redistribution of molecular and nanoparticle functionalities effectively on opposite sides of a single microparticle, giving rise to an architecture we term “antipodal” (Fig.\ref{fig1}C). With magnetic nanoparticles, the spontaneous asymmetric assembly provides emergent magnetic anisotropy. Theoretical modelling of these processes provides mechanistic insight into the origin and evolution of the experimentally observed asymmetry and anisotropy, and establishes a framework for controlling the emergent architecture.
\bigskip

Our multifunctional particles with internal magnetic anisotropy and antipodality enable magnetic steering \cite{zhou2021magnetically, vashist2026magnetically} and catalytic propulsion \cite{sengupta2013enzyme, meng2025micro, chen2024enzymatic} simultaneously without the need of multi-step fabrication typically required for magnetic dual-driving systems \cite{zhou2021magnetically, wu2022motion, villa2018cooperative, unruh2023remote, lu2016catalytic}. Exploiting the emergent magnetic anisotropy, we demonstrate gradient-free directed enzymatic propulsion (Fig.~\ref{fig1}E). We also demonstrate another application where dual-enzyme antipodal architecture enhances weak chemotaxis by a homogeneously abundant fuel (Fig.\ref{fig1}F).  We further demonstrate the versatility of the platform by incorporating various enzymes, nanoparticles, and photocatalytic components for dye removal, as well as encapsulating and releasing hydrophilic (carboplatin) and hydrophobic (curcumin) drugs. Together, our experimental and theoretical results establish a fully aqueous strategy for the spontaneous spatial organization of multiple functionalities within individual micromotors, opening new possibilities for designing microscale systems with integrated propulsion and directional control.

\begin{center}
   \begin{figure}[htbp] 
		\centering
		\includegraphics[width=0.9\textwidth]{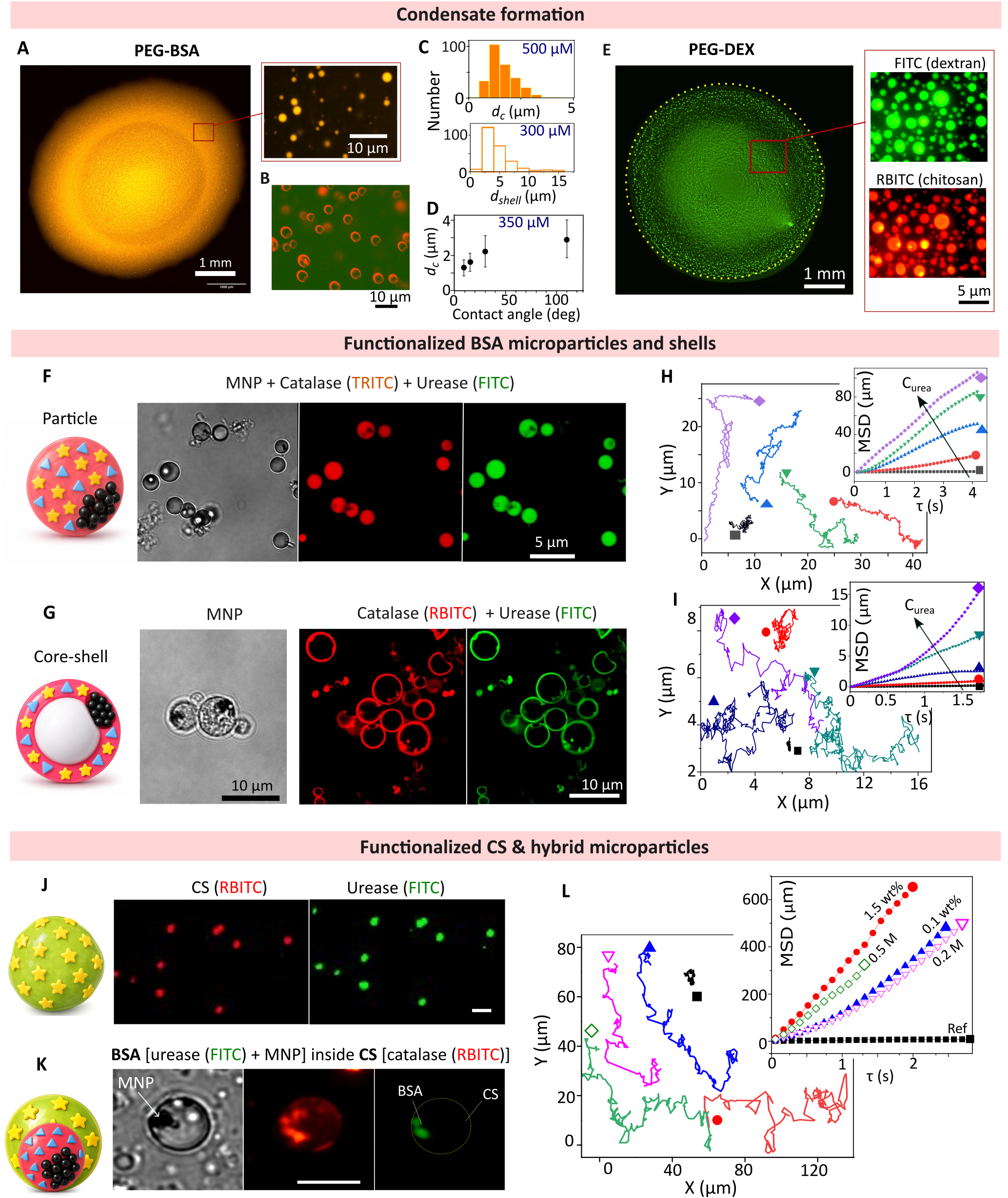}
		\caption{ \textbf{Microcondensate formation in evaporating droplet:}  \textbf{A} PEG-BSA aqueous droplet containing NH$_3$ (2.5~wt\%) formed numerous microcondensates upon evaporation. C$_{BSA}$ = 500 $\mu$M and C$_{PEG}$ = 1~wt\%. BSA was tagged with tetramethylrhodamine isothiocyanate (TRITC).  \textbf{B} Core-shells were formed at C$_{BSA}$ = 100 $\mu$M.  \textbf{C} Size distribution of the BSA condensates and shells. \textbf{D} Condensate size with contact angle measured on plasma-treated glass and PDMS surfaces (110$^\circ$). \textbf{E}  Evaporation of PEG--DEX droplets with chitosan (CS) forms stable DEX-CS microcondensates (inset). Initial $C_{\mathrm{PEG}}$ = 6 wt\%, $C_{\mathrm{DEX}}$ = 1~wt\% and $C_{\mathrm{CS}}$ = 1~wt\%. 
         \textbf{Activity of crosslinked condensates in solution bath:}  \textbf{F, G} BSA particles and shells with dual (catalase \& urease) or triple (magnetite nanoparticles (MNP), catalase, \& urease) functionality. The enzymes were tagged by TRITC, rhodamine-B-isothiocyanate (RBITC), or fluorescein isothiocyanate (FITC), as shown in the panel.  \textbf{H} Trajectories and corresponding MSD's of urease-loaded particles and shells (\textbf{I}) at urea concentrations, $C_{urea} = 0, 50, 100, 200, 400$ mM. Particle diameter: 2 $\mu$m. Shell diameter: 10 $\mu$m. \textbf{J} CS particles with catalase and urease. \textbf{K} CS particles containing catalase and small BSA particles functionalized with MNP and urease. \textbf{L} Trajectories and corresponding MSDs of BSA-particle-in-CS-particle construction, separately measured for a single particle for urea (open symbols) and H$_2$O$_2$ (filled symbols) for the enzymes urease and catalase, respectively. } \label{Fig2}
	\end{figure}
\end{center}

\section{Results }

In the following sections, we first describe the evaporative mechanism underlying condensate microdroplet formation and the asymmetric localization of components within the condensates. Then we demonstrate proof-of-concept applications of these active antipodal particles.

\subsection{Microcondensate formation in evaporating multiphase droplets}

We use two representative aqueous phase-separating systems: (i) polyethylene glycol (PEG)-bovine serum albumin (BSA) and (ii) PEG-dextran (DEX)-chitosan (CS). In each system, upon evaporation of the mixture droplet, liquid-liquid phase separation produces numerous micron-sized condensates of BSA or DEX near the droplet edge (Fig.~\ref{Fig2}).

To obtain a narrow size distribution, it is essential to suppress the coalescence of these condensates. While pure PEG-BSA droplet evaporation produced large clusters (Supplementary Information), the addition of ammonium hydroxide produced microcondensates of BSA by increasing the protein charge and reducing the interfacial tension, thereby suppressing condensate coalescence \cite{jambon2024phase}. This produced numerous condensates with typical diameters of $\sim 2~\mu$m (Fig.~\ref{Fig2}A). At lower BSA concentrations (typically $<300~\mu$M), however, the condensates appeared as shell-like structures (Fig.~\ref{Fig2}B), likely due to evaporation redistributing BSA toward the interface and a crowded PEG-rich interior becoming entropically unfavorable for BSA \cite{patel2022macromolecular}. These shell-like structures were larger, with diameters of approximately $6 \pm 2~\mu$m (Fig.~\ref{Fig2}D). With BSA concentration, the morphological transition into shells is gradual, with smaller concentrations producing thinner BSA shells.

During evaporation of pure PEG-DEX mixture droplets, condensate coalescence leads to a central accumulation of DEX \cite{Guo2021, Rai2024} (also in the Supplementary Information). However, by incorporating chitosan (CS), the PEG-DEX-CS system formed microcondensates, in which CS adsorbs at the PEG-DEX interface and stabilizes the condensates. Consequently, uniformly distributed condensates with a typical diameter of $d_c = 2.5 \pm 0.5~\mu$m were obtained (Fig.~\ref{Fig2}E).

In both systems, the condensate size almost doubles toward the droplet center, as larger condensates preferentially migrate weakly inward (Supplementary Information). Also, slightly smaller condensates are obtained at lower substrate contact angles (measured with respect to water), as shown in Fig.\ref{Fig2}D. This can be attributed to the enhanced evaporation rate, which drives the composition through the phase separation more rapidly, resulting in a higher condensate density and, consequently, smaller condensates.

\subsection{Microcondensates to multifunctional particles}

Phase-separating microcondensates encapsulate biomolecules by preferential partitioning  \cite{keating2012aqueous, ramos2022protein}.  For example, in a PEG-DEX polymer system, the DEX phase preferentially encapsulates proteins and nucleic acids \cite{sakuta2020self, Guo2021, Qi2024, hu2023liquid}, while phospholipids and nanoparticles tend to accumulate at the phase boundary \cite{keating2012aqueous, zhang2021engineering}.

Building on this behavior, functional components, including catalase, urease, magnetite (Fe$_3$O$_4$), and silver nanoparticles, are incorporated into microcondensates by simply dispersing them initially into the evaporating droplet. These components are spontaneously encapsulated within the microcondensates and core-shells without additional processing. 

Upon curing the condensates with glutaraldehyde (outside the evaporating droplet), a widely used cross-linker for biocompatible applications \cite{barbosa2013glutaraldehyde, ma2016biocompatible}, we obtain stable, functionalized BSA particles and shells, as well as CS particles. Representative images of them are shown in Fig.~\ref{Fig2}F$-$K. 

To evaluate enzymatic activity, the functionalized particles were introduced into a bath of solutions containing the corresponding substrates: H\textsubscript{2}O\textsubscript{2} for catalase and urea (NH\textsubscript{2}CONH\textsubscript{2}) for urease. The resulting reactions

\[  2H\textsubscript{2}O\textsubscript{2} \rightarrow 2H\textsubscript{2}O + O_2\]
\text{and}
\[ NH_2CONH_2 + H_2O \rightarrow CO_2 + 2NH_3 \]

generate local concentration gradients around the particles that drive self-diffusiophoretic propulsion \cite{wang2013small, zhao2018powering, patino2018influence, golestanian2005propulsion}. This active motion becomes randomized at longer times due to rotational diffusion. \medskip

The enhanced motion of particles and shells is evident from the trajectories and the corresponding mean-squared displacement (MSD) as plotted in Fig.~\ref{Fig2}H~\&~I for the encapsulated urease enzyme. When catalase was encapsulated, activity was observed in H$_2$O$_2$ solution.

We can also fabricate hierarchical hybrid architectures of BSA and CS or vice versa. For example, in Fig.~\ref{Fig2}K, pre-formed functional BSA particles were suspended in PEG-DEX-CS droplets and encapsulated within DEX--CS condensates. Magnetite nanoparticles (MNP) were incorporated simultaneously, yielding hybrid particles in which urease + magnetite-functionalized BSA particles were within a catalase-functionalized CS matrix. These combinations can be varied appropriately. BSA and MNP are asymmetrically located within the CS. Although catalase is distributed throughout the CS matrix, a slight enrichment is observed near BSA-containing regions (Fig.~\ref{Fig2}K).
The activity of these particles, tested sequentially in H$_2$O$_2$ and urea solutions, is shown in Fig.~\ref{Fig2}L, demonstrating stable enzymatic activity.

 Although glutaraldehyde is known to decrease enzyme activity, as shown in  $\beta$-galactosidase \cite{chen2013The} and urease \cite{LiangUrease}, the high enzyme loading retained strong particle activity. The encapsulation efficiencies of different species varied between DEX and BSA condensates, as quantified using the bicinchoninic acid protein assay (Supplementary Information). For example, urease and catalase exhibited encapsulation efficiencies of 66\% and 72\%, respectively, in DEX-CS condensates. In BSA condensates, they showed efficiencies of 87\% and 85\%, respectively. The encapsulation efficiency of MNP and BSA in CS particles was about 70\%, obtained from particle images (Supplementary Information).

\begin{center}
\begin{figure}[htbp] 
\centering
\includegraphics[width= 0.9\textwidth]{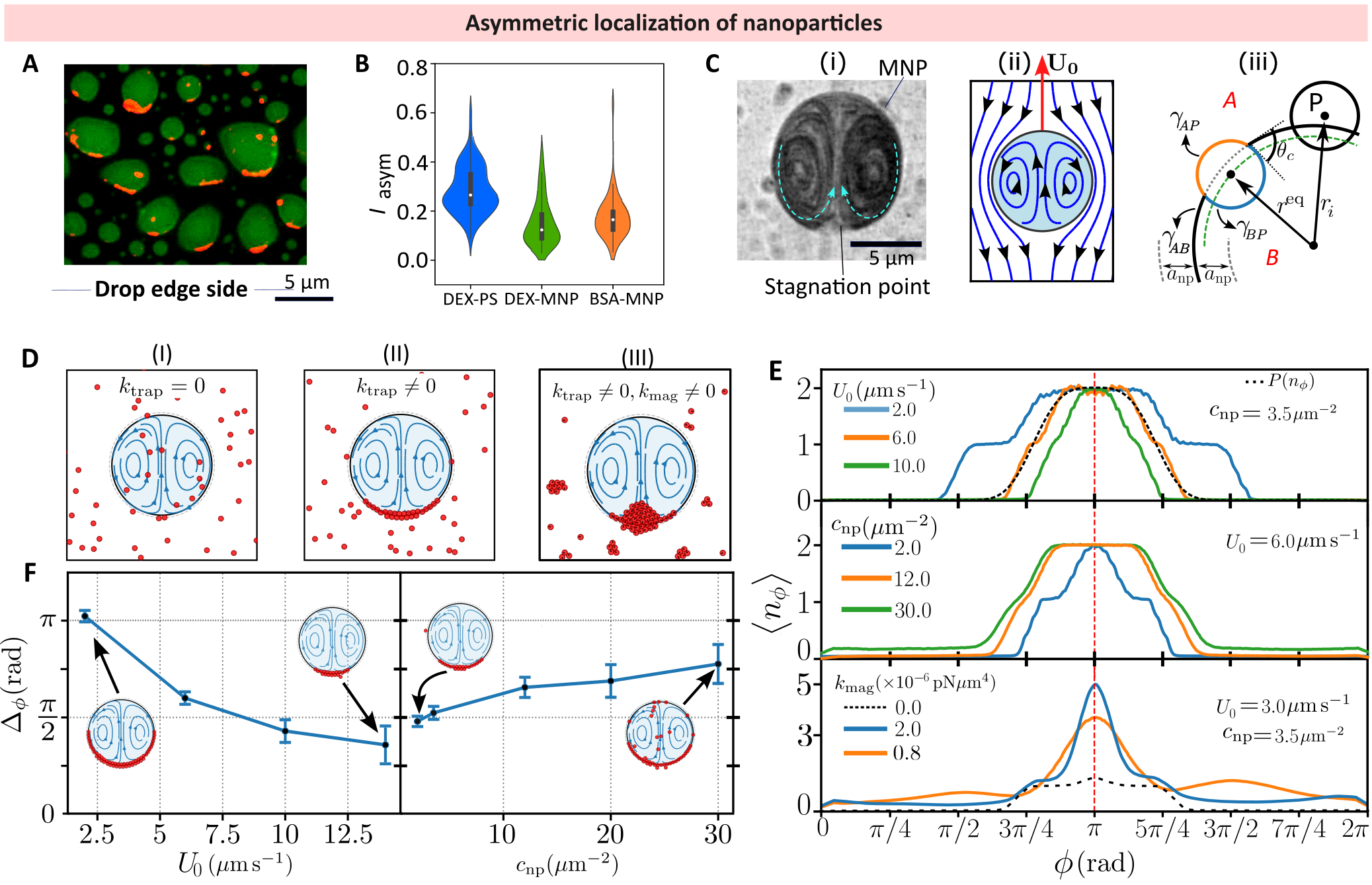}
\caption{ \textbf{Anisotropic internal clustering and crescent
formation inside the evaporating droplet:} \textbf{A.} Fluorescent polystyrene nanoparticles (100
nm, red) accumulate within DEX condensates, with the stagnation region oriented
toward the droplet edge (Initial $C_{\mathrm{PEG}}$ = 6 wt\% and
$C_{\mathrm{DEX}}$ = 1~wt\%).  \textbf{B.} Distribution of the asymmetry factor
$I_{\rm asym}$ in dextran--polystyrene (DEX--PS), dextran--magnetite
nanoparticle (DEX--MNP), and BSA--MNP condensates.  $I_{\rm asym} < 0.5$
implies that the nanoparticle region is less than half of the condensate area.
Data from 80 to 100 condensates.  \textbf{C.} Circulatory flow inside the condensate in the particle body frame (i) and the corresponding Marangoni flow field used in the theory (ii), showing both intra- and extra-condensate flows.  The red upward arrow denotes the condensate self-propulsion velocity, $U_0$, towards the droplet center.  Panel (iii) schematically shows the interfacial components and nanoparticle configuration used to define the
interfacial energy and corresponding trapping force, $\Fb_i^{\rm trap}$.
\textbf{D.} Steady-state simulation snapshots of non-magnetic nanoparticles without and with interfacial-tension-mediated retention (panels I and II,
respectively), and a snapshot of magnetic nanoparticles after the crescent
collapses into a bead (panel III); $k_{\rm{trap}}=1.98 ~\rm{pN} \mu \rm{m}^{-1}, \theta_c=\pi/2$ for panel II and $k_{\rm{trap}}=1.4~\rm{pN} \mu m^{-1}, \theta_c\simeq 0.9\pi/2$ 
 for panel III ; $\gamma_{AB}=1 ~\rm{pN} \mu \rm{m}^{-1}$ \cite{Cristinasurfacetension2006} for both.  Arrowheads on the
red nanoparticles indicate the directions of their magnetic moments.
\textbf{E.} Time-averaged surface occupancy, $\langle n_\phi \rangle$, obtained
from simulations, as a function of the interfacial angle $\phi$ for varying self-propulsion velocities $U_0$ (top), nanoparticle concentrations $c_{\rm
np}$ (middle), and magnetic moments of magnetic nanoparticles (bottom).
\textbf{F. Tunable crescents:} Theoretical predictions demonstrating control
over the angular width of the crescent through variations in the
self-propulsion velocity, $U_0$, and nanoparticle concentration, $c_{\rm
np}$.}

\label{Fig3}

\end{figure}
\end{center}

\subsection{Marangoni-driven asymmetric organization of nanoparticles} \label{Masymm}

During evaporation, surface-tension gradients along the droplet-air interface induce large-scale circulation in the surrounding fluid, as illustrated in Fig.~\ref{fig1}B. The surface-tension gradients are due to the compositional gradients in the droplet arising from the phase separation near the edge. Simultaneously, the compositional gradients around the microcondensates generate Marangoni stresses at the condensate-liquid interface that drive recirculating flows inside the condensates.  This is the reason for the condensate migration towards the droplet center \cite{Guo2021, May2022, Rai2024} (with velocity $\textbf{U}_0$). The flows inside and around the condensates play a central role in the capture and self-organization of nanoparticles inside the microcondensates, as discussed below.

A key feature of our method is the spontaneous clustering and asymmetric
localization of solid constituents (nanoparticles) within the phase-separated
condensates (Fig.~\ref{Fig2}\& \ref{Fig3}A). Experimentally, to quantify the spatial partitioning of nanoparticles within the condensates, we define an asymmetry index,

\begin{equation}
I_{\rm asym}=\frac{A_{np}}{A_{np}+A_c},
\end{equation}

where $A_{np}$ and $A_{c}$ are the projected areas of the nanoparticle-rich and
nanoparticle-poor regions, respectively, of the condensates. A value of $I_{\rm asym}$ = 0.5
corresponds to an ideal Janus-like partition with equal areas occupied by the
two domains, whereas values approaching 1 indicate an increasing dominance of
the nanoparticle-rich region. The distribution of $I_{\rm asym}$ shown in
Fig.~\ref{Fig3}B demonstrates that the condensates exhibit Janus-like
internal architecture with moderate condensate-to-condensate variation in the relative size of the nanoparticle-rich domain.

We can observe two important characteristics of the
condensate-nanoparticle assembly: 
\begin{enumerate}[label=(\roman*)]
    \item The accumulated nanoparticles form an extended crescent-shaped cluster on the upstream hemisphere (opposite to $\Ub_0$) of the condensate [Fig.~\ref{Fig3}A]. The cluster breaks the condensate's head-tail symmetry, creating the Janus-like architecture.
    
    \item For magnetic nanoparticles, the cluster spontaneously collapses into a compact anisotropic bead [Fig.~\ref{Fig2}F\& G]. The shape anisotropy gives rise to strong field-induced apolar alignment, while its intrinsic magnetic moment provides weak polar alignment, together enabling versatile control over the condensate's single-particle dynamics (vide infra). 
\end{enumerate}

\textbf{\textit{Origin of crescents and beads.}} We now establish that the combined
effects of previously overlooked surface-tension-mediated nanoparticle
adsorption and retention, together with flows within and around the condensate,
underlie the formation of crescents and beads. Fig.~\ref{Fig3}C (i) shows
the circulatory flow in the particle frame, measured experimentally within the
condensate [Supplementary Video~1]. It is consistent with the theoretical Marangoni flow field, extending both inside and outside the condensate and generated by interfacial surface-tension gradients [Fig.~\ref{Fig3}C (ii)].

The Marangoni flows break the front-back symmetry of the condensate with
respect to $\Ub_0$, inducing a persistent advective flux along the interface
towards the downstream pole \cite{zhang2021engineering}. However, as shown by
our hydrodynamic simulations in Fig.~\ref{Fig3}D(I), this flow alone is
insufficient to form a crescent or bead at the pole. Most nanoparticles follow
the streamlines and are swept away from the interface, both inside and outside
the condensate, while a small fraction near the interior surface is
periodically brought to the interface by the circulatory flow [Fig.~\ref{Fig3}C(ii), Supplementary Video~2, and Supplementary Information].

\textbf{\textit{Surface-tension-mediated interface retention.}} While
Marangoni-driven advection alone is insufficient to capture the observed
nanoparticle dynamics, the complete dynamics additionally involve interfacial
thermodynamics and interparticle forces. The interfacial contribution is
described by \cite{Binks_Horozov_2006} $G(r_i) = \gamma_{AP}A_{AP}(r_i) + \gamma_{BP}A_{BP}(r_i) +
\gamma_{AB}A_{AB}(r_i)$, where $\gamma_{\alpha\beta}$ and $A_{\alpha\beta}$
denote the interfacial tension and area between phases $\alpha$ and $\beta$,
respectively; $A$, $B$, and $P$ denote the phases outside and inside the
condensate and the nanoparticle material, respectively [Fig.~\ref{Fig3}C(iii)].
As the interfacial areas depend on the radial position $r_i$ of nanoparticle
$i$ relative to the condensate center, $G$ is minimized at a preferred position
$r^{\rm eq}$, giving rise to the interfacial trapping force,

\begin{align}
\Fb_i^{\rm trap}=-k_{\rm trap}(r_i-r^{\rm eq})\,\rhat_i,
\label{eq:ftrap}
\end{align}

where $k_{\rm trap}=2\pi\gamma_{AB}$ [Supplementary Information for full derivation]. The force tends to retain the nanoparticle near $r^{\rm eq}$ and acts only while the nanoparticle remains in contact with the interface, i.e., while its centre lies within the nanoparticle radius, $a_{\rm np}$, of the interface on either side.

Including $\Fb^{\rm trap}_i$ in our simulations reveals the crucial role of
interfacial trapping: nanoparticles now adsorb onto and remain at the interface,
counteracting the outward component of the Marangoni flow that would otherwise
sweep them away [see Supplementary Video~3]. Incoming nanoparticles
are thus captured at the upstream interface by the combined action of the
Marangoni flow and $\Fb^{\rm trap}_i$, then swept along the interface towards
the downstream pole, producing the crescent observed for non-magnetic
nanoparticles in Fig.~\ref{Fig3}D(II).

For magnetic nanoparticles, an initial crescent forms with nearly parallel dipole moments $\pb_i$, minimizing the magnetic energy. As the crescent widens, however, its curvature frustrates this alignment, causing the chain to break
and rapidly collapse into a compact bead-like structure [Fig.~\ref{Fig3}D(III) and Supplementary Video 4]. The resulting structure allows locally parallel intra-chain and staggered inter-chain arrangements, together with near-loop arrangements across the bead, thereby lowering the magnetic energy globally.

Remarkably, the equilibrium contact angle $\theta_c\simeq0.9\pi/2<\pi/2$
[Fig.~\ref{Fig3}D(III)] gives $r^{\rm eq}=r_c-a_{\rm np}\cos\theta_c<r_c$, where
$r_c$ is the condensate radius, causing the bead to protrude substantially into
the condensate interior, as observed experimentally. While the edge nanoparticles remain retained at the interface by $\Fb_i^{\rm trap}$, the interior nanoparticles are anchored to them by magnetic attraction alone.
Since the interior flow strengthens towards the condensate center, sufficiently large beads are fragmented into the interior in our simulations. Occasionally, we observe this behavior in the experiments. We speculate that, in the experiments, sufficiently large beads contact the substrate due to their size and weight and are thereby stabilized by substrate interactions. The effects of the substrate on Marangoni flows and bead stabilization lie beyond the scope of the present study and will be
addressed elsewhere.

\textbf{\textit{Tunability of crescents and beads.}} Fig.~\ref{Fig3}E (top and middle) shows the simulated time-averaged interfacial occupancy $\langle n_\phi\rangle$ for nonmagnetic nanoparticles as a function of angular position $\phi$. Most profiles exhibit a plateaued peak at $\langle n_\phi\rangle=2$, indicating a
double-layered crescent core with a thinner periphery. The core widens with decreasing $U_0$ at fixed nanoparticle concentration $c_{\rm np}$ (top) or increasing $c_{\rm np}$ at fixed $U_0$ (middle). For small $U_0$ or $c_{\rm np}$, the prominent steps at $\langle n_\phi\rangle=2$ and $1$ indicate a sharply defined core--periphery boundary and a single-layered periphery, respectively, whereas smoother profiles indicate a rapidly fluctuating boundary. The plateaued profiles closely follow a \emph{flat-top Gaussian} (dashed curve, top). \cite{laserssvelto2010}

Nonzero magnetic moments qualitatively change the $\langle n_\phi\rangle$ vs.~$\phi$ profiles (bottom): the peak becomes rounded and Gaussian-like, while its height increases markedly and its width decreases. This indicates that most nanoparticles in the cluster concentrate within its core, consistent with the collapse of the crescent into a bead [Fig.~\ref{Fig3}D(III)].

Fig.~\ref{Fig3}F shows the widths of the $\langle n_\phi\rangle$ vs.~$\phi$ profiles for nonmagnetic nanoparticles, extracted from Fig.~\ref{Fig3}E (top and middle), as functions of $U_0$ and $c_{\rm np}$. Tuning the two experimentally controllable parameters enables substantial control over the crescent width, with variations up to $\sim300\%$. The peak height in
Fig.~\ref{Fig3}E (bottom) likewise indicates significant tunability of the magnetic bead size with the nanoparticle magnetic moment.

Experimentally, $U_0$ can be tuned by spatially varying the evaporation rate. A faster evaporation rate towards the droplet edge generates stronger concentration gradients of the phase components and an enhanced overall Marangoni effect. Under our ambient evaporation conditions and in the presence of surface-active components such as CS or ammonia, we have a relatively low $U_0$, typically \(<10~\mu\text{m/s}\), compared with the pure system \cite{May2022, Rai2024}.

\begin{figure} [htbp]
\centering
\includegraphics[width=0.90\linewidth]{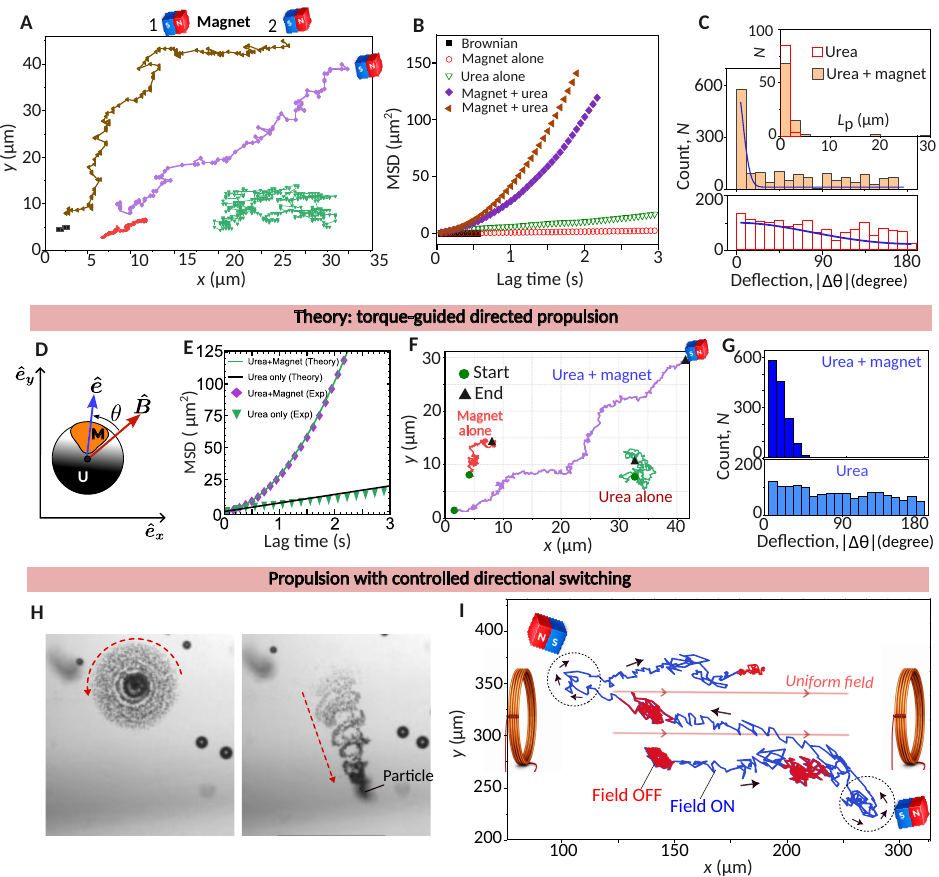}

\caption{\textbf{Dual-functionalized condensate:}  \textbf{A} Trajectories and \textbf{B} corresponding MSDs of a BSA particle containing urease and a magnetite cluster under different conditions (shown in \textbf{A}). In homogeneous urea (0.5 M), catalytic propulsion amplifies the magnetophoresis. Moving the neodymium magnet from position 1 to 2 redirects the field gradient
and the trajectory.  Trajectory duration: 25~s. \textbf{C} The distribution of the turning angle $\Delta\theta$ (see text) obtained from corresponding directed trajectories shown in \textbf{A}. Small $\Delta\theta$ indicates directionality.  The blue curve on the histograms is the fit of Eq.\ref{eq_vonMises}. Also, the overall persistence length $L_p$ increased with dual driving (inset). Particle diameter: 2.6 $\mu$m. The magnet position: 2 cm from the solution.  \textbf{Condensate dynamics from theory:} \textbf{D} Schematic of the dual-functionalized condensate model used in simulations; $M$ and $U$ denote the antipodally placed magnetic nanoparticle cluster and urea-coated surface region, respectively.  \textbf{E} Theoretical fit to the experimental MSD, yielding the diffusion constants $\tilde{D}_{\rm tr,rot}$ and
magnetic alignment strength $\alpha_{\rm pol}$. \textbf{F} Simulated trajectories corresponding to the red, green, and purple trajectories in
\textbf{A}. \textbf{G} Distribution of turning angles $\Delta\theta$ from the urea+magnet and urea-alone trajectories in \textbf{F}. \textbf{Controlled directional switching and orientation-retention:} \textbf{H} BSA particle with a magnetite cluster and Ag nanoparticles in H$_2$O$_2$ solution rotates like a 'firecracker wheel' by bubble propulsion, but moves in a direction when the uniform field is on.  \textbf{I} A brief magnetic-field gradient flips the particle, which otherwise remains stably oriented by the apolar anisotropic torque.  }

\label{fig4}
\end{figure}

\subsection{Application: gradient-free directional propulsion with controllable
directional switching} \label{secApp1}

Magnetic guiding is an excellent strategy for drug delivery at known target locations \cite{tang2026enzymatic, zhang2021dual, naikwadi2024self, chen2025roadmap}. Magnetic manipulation combined with catalytic propulsion has emerged
as an efficient strategy for microscale navigation, as demonstrated in systems
such as magnetic particles and partially catalytic discs
\cite{zhou2021magnetically, wu2022motion, luo2020enhanced,
villa2018cooperative, unruh2023remote, lu2016catalytic}. In most of these approaches, magnetic and catalytic functionalities, as well as asymmetries, are introduced through multi-step fabrication processes, unlike our system where these properties emerge spontaneously. We demonstrate three major aspects of directional driving and control.

\begin{enumerate}[label=(\roman*)]
\item In a uniform magnetic field, due to the magnetic inclusion, BSA particles experience a torque \cite{suwa2023applications}, 

\begin{equation}
\mathbf{T}_{mag} = p^{\rm{net}}B(\ehat\times\Bhat) + \dfrac{B^2V\,\Delta\chi}{\mu_0} (\ehat\cdot\Bhat)(\ehat\times\Bhat),
\label{eq_Tm}
\end{equation}

where the first term on the right-hand side is the polar contribution
arising from the net magnetic moment of the condensate's magnetic bead,
$\pb^{\rm net}=p^{\rm net}\ehat$, while the second term is the apolar
contribution arising from susceptibility anisotropy,
$\Delta\chi=\chi_{\parallel}-\chi_{\perp}$, associated with the shape
anisotropy of the magnetic inclusion. Here, $\Bb=B\Bhat$ is the applied
magnetic field, $\mu_0$ is the vacuum permeability, and $V$ is the
magnetic-inclusion volume. The magnetic torque aligns the body-fixed
magnetic axis and suppresses rotational fluctuations, thereby directing the chemically driven motion in a uniform field, without requiring an external chemical gradient (vide infra - iii).

\item In a nonuniform magnetic field, on top of the torque, the magnetic bead provides an additional isotropic magnetophoretic force, 

\begin{equation}
\mathbf{F}_{\rm mag} = \nabla(\pb^{\rm{net}}.\,\,\Bb) + \dfrac{\chi V}{2\mu_0}\nabla B^2 
\label{eq_Fm}
\end{equation}

The magnetic moment moves the particles directionally under small magnetic-field gradients, while enzyme catalysis amplifies the motion. 

\item The chemical functionality is localized predominantly on the side opposite to the magnetic bead, giving effectively an antipodal architecture (see Fig.\ref{fig1}C or \ref{fig4}D). This spontaneous asymmetry plays an important role in coupling magnetic orientation to chemical propulsion. It enables motion parallel or antiparallel to a uniform magnetic field by chemical propulsion when the particle's magnetic axis is oriented along the field. Without antipodal architecture, the chemical propulsion direction may be misaligned with the magnetic axis, causing the particle to propel along directions other than the magnetic-field axis. We hypothesize that this antipodal organization arises because regions not occupied by the magnetic nanoparticle cluster remain accessible for enzyme incorporation.

\end{enumerate}
  
\textbf{\textit{Field-oriented chemical amplification of magnetophoresis}:}

Here, we use BSA particles containing urease and MNP. In the absence of a magnetic field but in the presence of homogeneously distributed urea, the particles exhibited strong random active motion. Without urea, when only the magnetic gradient was applied (using a permanent magnet), the force $F_{mag}$ caused the particles to show weak directional motion toward the external magnet, as inferred from the trajectories in Fig.~\ref{fig4}A.

In contrast, when both the magnetic field gradient and urea were present, the particles exhibited increased motion, powered by the uniformly available urea substrate, without losing directionality towards the magnet. Dynamical steering is possible by moving the magnet position as shown in Fig.\ref{fig4}A. 

The MSDs show enhanced particle motion (Fig.\ref{fig4}B). Fitting the MSDs (see Methods) yielded particle velocities \cite{PREL052602} of 0, 0.2 $\pm$ 0.004, 0.7 $\pm$ 0.01, and 6.0 $\pm$ 0.08 $\mu$m/s for Brownian motion alone, magnetic field alone, urea alone, and magnetic field + urea, respectively. These trajectory measurements were obtained for a single particle to enable direct comparison across conditions. The above values show that the particle velocity is amplified by up to 30 times in dual powering compared to magnetophoresis. 

Applying a magnetic field substantially reduced the rotational fluctuations of the particles, as reflected by the narrower distribution of deflection angles ($\Delta\theta$ - also see Fig.1E) compared to the case of urea alone, as shown in Fig.\ref{fig4}C. Also, the persistence length $L_p$ (distance between each consecutive deflection) increased in the dual driving.

We establish the chemical amplification of magnetophoresis theoretically by modelling the dynamics of the doubly functionalized condensate (see Methods), with the resulting
trajectories and analysis shown in Fig.~\ref{fig4}D--G. As in experiments, a weak magnetic field alone ($B\ne0$, with small $F^{\rm mag}$) produces a random
walk with a small average magnetophoretic drift, while chemical driving alone ($v_0\ne0$) yields a \emph{persistent random walk} with negligible directed motion over the observation timescale (red and green trajectories, Fig.~\ref{fig4}F). Activating both functionalities ($B,v_0\ne0$) instead
produces a strongly amplified magnetophoretic trajectory (purple), despite the weak magnetic force. The field thus primarily orients rather than translates the condensate, suppressing rotational diffusion and reducing deflection-angle fluctuations, as shown in Fig.~\ref{fig4}G.

\textbf{ \textit{Controlled directional switching and orientation-retention in uniform fields:}}

Magnetophoresis requires a magnetic field gradient, which, the induced-magnetization contribution, decays as $r^{-4}$. Consequently, the magnetic force scales as $r^{-7}$. This rapid decay necessitates very large magnetic fields to drive motion deep inside the body. In contrast, a uniform magnetic field cannot generate translational motion, but it can align particles possessing magnetic shape anisotropy when combined with chemical propulsion.  

To demonstrate directional motion without magnetic field gradients, we produced a homogeneous magnetic field using a Helmholtz coil. We used BSA particles containing silver and magnetite nanoparticles, suspended in a homogeneous H$_2$O$_2$ solution (1.5 wt\%), where catalytic decomposition generated $\mathrm{O}_2$ bubbles. Without the fuel (H$_2$O$_2$), these particles showed no net motion but Brownian fluctuations under the uniform field.

With the fuel but without a magnetic field, particles exhibited rapid rotational motion with little translational motion due to the angled bubble ejection (Fig.~\ref{fig4}H and Supplementary Video 5). Upon applying a uniform magnetic field, the particles began to translate directionally. Here, the magnetic torque suppresses the particle's strong active rotation and stabilizes it, so the motion remains biased in a single direction (Supplementary Video 5).  

Unlike a magnetic field gradient, a uniform magnetic field exerts a torque
but no net translational force on the magnetite nanoparticle cluster
embedded within the BSA particles. The resulting torque can orient the
particles with their body-fixed magnetic axis either parallel or
antiparallel to the field. The orientation can be switched by briefly
applying a magnetic field gradient, which exerts a force on the weak net
magnetic dipole moment ($\pb^{\rm net}$) of the magnetic bead inclusion,
as shown in Fig.~\ref{fig4}I. Once flipped, the particle continues to
propel in the opposite direction while the uniform field is maintained,
and returns to its original direction only upon a subsequent flipping
event. The persistence of both orientations under the same applied
uniform field indicates that the apolar torque associated with
susceptibility anisotropy is sufficiently strong to stabilize the two
opposite orientations despite the presence of a weak polar contribution.
Similarly, rotating the uniform field direction also enables dynamic control of its direction (Supplementary Information).
\medskip

To quantify the suppression of directional fluctuations by the magnetic field, we analyzed the distribution of the absolute turning angles ($|\Delta\theta|$) between successive trajectory segments (Fig.~\ref{fig4}C). The measured distributions were fitted with a folded Gaussian function

\begin{equation}\label{eq_vonMises}
    P(|\Delta\theta|)=A
\exp\left( -\frac{|\Delta\theta|^2}{2\sigma_\theta^2}\right),
\qquad 0\le|\Delta\theta|\le\pi.
\end{equation}

For small angular deviations from the field direction, the rotational Langevin equation gives steady-state variance $\sigma_\theta^{2} = \tilde{D}_{\rm{rot}}/\alpha$, with $\alpha = \tilde{\mu}_{\rm{rot}}(\alpha_{\rm{pol}} + \alpha_{\rm{apol}} )$ as an effective angular alignment rate and $\tilde{D}_{\rm{rot}}$ as the rotational diffusion coefficient of the BSA particles (Methods). Larger $\alpha$ values correspond to stronger directional persistence and a narrower turning-angle distribution. 

The fitted values decreased from $\sigma_\theta^{2}$=3.4$\pm$0.8 in the absence of the magnetic field to $\sigma_\theta^{2}$=0.017$\pm$0.006 with the magnetic field. The latter corresponds to $\alpha>>\tilde{D}_{\rm{rot}}$, indicating a substantial suppression of random diffusive fluctuations.

Using the measured $\alpha$ and considering the rotational drag of a particle volume $1\mu\mathrm{m}^3$, we estimate an effective susceptibility anisotropy of order $10^{-3}$.  We can estimate the corresponding torque $T_{mag}$ as $\sim10^{-18}\,\mathrm{Nm}$ for $B \sim 10~\mathrm{mT}$, the typical field strength used in our experiments. This torque corresponds to an orientational energy scale \(T_{\rm mag}\Delta\theta\) that exceeds \(k_BT\), consistent with the strong magnetic response of the particles. 

\bigskip
\bigskip

\begin{figure} [htbp]
\centering
\includegraphics[width=0.90\linewidth]{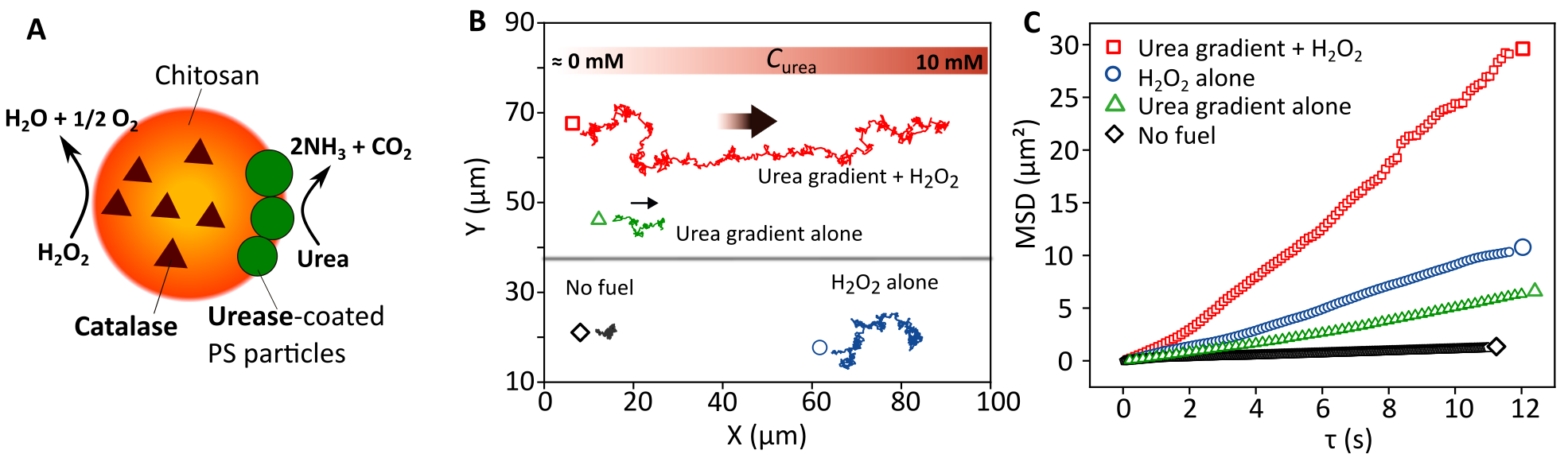}

\caption{\textbf{Amplification of chemotaxis:}  (Also refer to Fig.\ref{fig1}C(ii) \& F).  \textbf{A} Schematic of the chitosan micromotor incorporating catalase and urease, with urease immobilized on polystyrene nanoparticles to achieve asymmetric localization. \textbf{B, C} Trajectories and MSD demonstrating enhanced motion driven by the H$_2$O$_2$--catalase reaction and steered by the urea--urease reaction. A weak urea gradient induces only weak migration ($\triangle$), whereas the additional presence of homogeneously distributed H$_2$O$_2$ enhances motion along the chemotactic direction ($\square$) through the antipodal arrangement of the enzymes. The velocities obtained from the MSDs are: 0 ($\diamond$), 0.03 ($\triangle$), 0.06 ($\bigcirc$), and 0.5 ($\square$) $\mu$m/s. The markers show the starting point of the trajectories. Micromotor diameter: 3 $\mu$m; trajectory duration: 200 s.  }

\label{fig4chem}
\end{figure}

\subsection{Application: Dual-enzyme amplification of weak chemotaxis}

Enzymatic micromotors can migrate along concentration gradients of their substrates \cite{dey2015micromotors, arque2019intrinsic, ghosh2023dual}. Although this phenomenon can be exploited for local targeted delivery, unfortunately, in natural environments, such as those surrounding tumour tissues, these gradients may be too weak to generate strong directional transport. This constraint can be overcome by a dual-enzyme strategy that separates steering from propulsion \cite{ghosh2023dual}, where the antipodality has an important role in coupling the strong propulsion to the steering along the weak gradient. We show that a weak urea gradient generates a self-diffusiophoretic torque \cite{PopescuNanolett, vinze2021motion} that steers the micromotors towards higher urea concentrations, while the spatially uniform presence of H$_2$O$_2$ sustains strong propulsion through the catalase reaction. This combination produces enhanced directional transport, characterized by long trajectories directed towards regions of higher urea concentration (Fig.\ref{fig4chem}).

We achieve this functionality by immobilizing urease on polystyrene (PS) particles and asymmetrically localizing these particles within chitosan microparticles. Catalase is simultaneously incorporated into the chitosan matrix. Both functionalities (catalase and urease-coated PS) were incorporated by mixing them into the evaporating droplet in a single step. The resulting antipodal arrangement spatially organizes the two enzymatic functionalities so that catalase-driven propulsion is aligned with the urease-mediated steering direction. Thus, a weak substrate (urea) gradient can guide strongly propelled micromotors towards its regions of higher concentration, using another abundant, homogeneously present substrate fuel (H$_2$O$_2$). Without the gradient but with H$_2$O$_2$ alone, the motion is strong but not directed. Importantly, without antipodality, the motion will be directional but not necessarily towards the weak gradient, affecting the concept of directed delivery.  

Dual-enzyme steering and propulsion have been demonstrated using peptide nanotubes as micromotor chassis \cite{ghosh2023dual}, although the underlying mechanism remained unexplored. Our findings suggest that enzyme coating may have produced an asymmetric, potentially antipodal distribution of the two enzymes, coupling gradient-dependent steering to substrate-driven propulsion. While this hypothesis remains to be tested in their system, our results highlight spontaneous spatial organization as a possible mechanism for dual-enzyme micromotor operation.

\subsection{Drug encapsulation and release}

Protein and polymer particles are promising biocompatible drug carriers \cite{lee2026benchmarking, roberts2020complex}. We demonstrated the proof-of-concept encapsulation of two model drugs, carboplatin, an anti-cancer drug, and curcumin. Especially hydrophobic drugs, such as curcumin, are difficult to deliver due to their poor availability in aqueous environments. At alkaline pH, curcumin's phenolic hydroxyl groups deprotonate, turning it into a negatively charged, water-soluble compound \cite{wang2021mechanism, chen2023co}.    
 
\begin{figure} [htbp]
    \centering
    \includegraphics[width=0.6\linewidth]{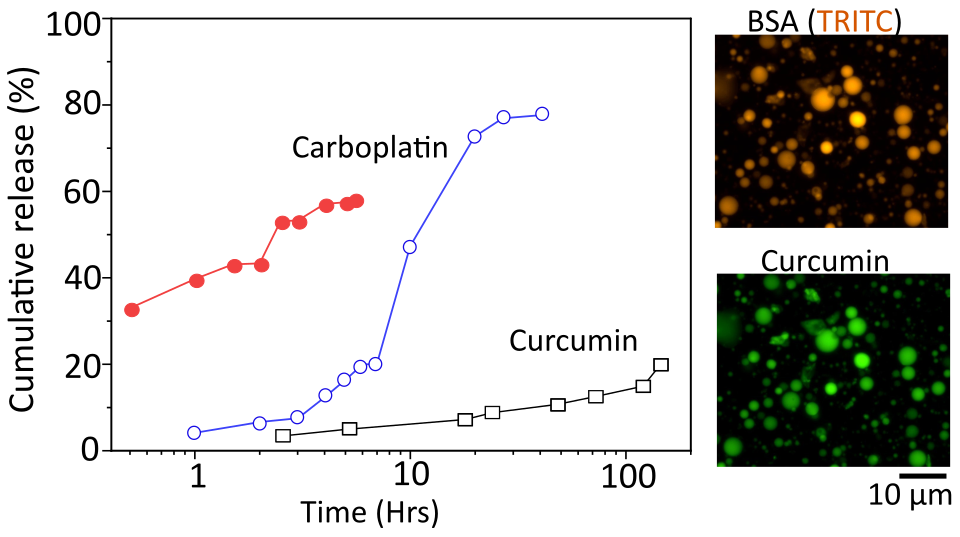}
   \caption{\textbf{Drug encapsulation and release}: The drug release measured by UV absorption spectra in the suspended aqueous solution for two encapsulated drugs, carboplatin (hydrophilic) and curcumin (hydrophobic) in chitosan (filled circle) and BSA (open circle and square) particles. Curcumin has a slower release rate than carboplatin. (Images): Curcumin shows its natural fluorescence (near 524 nm).   }
    \label{figDrug}
\end{figure}

BSA is widely known as a carrier for hydrophilic and hydrophobic drugs \cite{yu2017new, chen2023co}.  We observed that curcumin release from BSA was slow and attained below 15\% by 100 hrs (Fig.~\ref{figDrug} and Supplementary Information), possibly due to interactions based on hydrophobic effects between curcumin and BSA \cite{wang2021mechanism}. When the drug was hydrophilic (carboplatin in CS), we observed a release of approximately 40\% over 3 hrs. An increased and sustained release was observed for carboplatin from BSA matrix. This shows that these particle matrices can encapsulate drug molecules, whose spatial availability may be controlled by other functional molecules and MNPs. Hierarchical hybrid particle-in-particle architectures, for example, are attractive for gastrointestinal drug delivery, where carriers must withstand and traverse multiple harsh chemical environments \cite{zhang2025advanced, abbaspourrad2013controlling}.

\section{Discussion} 

To demonstrate generality, we tested the approach on several protein--polymer and polymer--polymer LLPS systems, including PEG--lysozyme, PEG--glucose oxidase, PEG--urease, and PEG--catalase mixtures. Except for HRP, PEG-rich condensates initially formed near the contact line in all cases. PEG--alginate (ALG) mixtures were also used to produce functionalized alginate particles (Supplementary Information).

The multifunctionality allows various other possibilities. For example, we fabricated BSA particles encapsulating both magnetite nanoparticles and photocatalytic graphitic carbon nitride (GCN) particles. Under UV illumination, GCN particles generate reactive oxygen species through interactions with water and dissolved oxygen, thereby degrading organic dyes \cite{zheng2025recent}. We demonstrated this using methylene blue as a model cationic dye and observed a reduction in dye concentration upon UV exposure. Owing to the incorporated magnetite nanoparticles, the BSA particles can be readily retrieved magnetically after the treatment process (Supplementary Information). 

Finally, our approach is potentially scalable by increasing the total contact-line length, for example, by using elongated droplets on hydrophilic stripes. Another method we demonstrate is to use a hydrophilic mesh that forms a thin suspended layer of solution. Evaporation from two air–liquid interfaces produces numerous microcondensates (Supplementary Information). In the latter configuration, one should explore methods to generate Marangoni flows, for example, by inducing concentration gradients, facilitating nanoparticle encapsulation. 

Several other directions emerge from this work. Our present theory predicts control over the structure that generates magnetic anisotropy. Future work will enable control over the resulting magnetic susceptibility anisotropy and may link it to the evaporation process and substrate effects. These aspects highlight the scalability and tunability of our approach while opening opportunities for further experimental and theoretical investigation.

\section{Conclusion}
We have demonstrated an evaporation-driven liquid–liquid phase separation strategy for engineering multifunctional protein- and polymer-based micromotors with emergent properties. Our approach enables the incorporation of diverse functional components, including enzymes, drugs, nucleic acids, and nanoparticles, in a single evaporation step. The simultaneous encapsulation of molecular and nanoparticle entities, and the spontaneous asymmetric localization of nanoparticles, allow an effective antipodal arrangement, in which distinct functionalities are localized on opposite sides of a micromotor. Magnetic and enzyme-functionalized nanoparticles allow broad antipodal architectures, while the anisotropic organization of magnetic nanoparticles introduces magnetic anisotropy. 

Through theoretical modelling, we show that the coupling of evaporation-driven Marangoni flows and interfacial dynamics underlies the formation of these spatially organized architectures. Thus, these complex features spontaneously emerge, which is otherwise challenging to achieve through conventional step-by-step fabrication approaches. The mechanistic understanding provides a basis for rationally tuning the spatial distribution of functional components and the resulting micromotor properties.

Exploiting the antipodal architecture, we demonstrate two modes of coordinated actuation. First, coupling chemical propulsion with magnetic torque enables directional transport in uniform chemical and magnetic fields. Second, antipodal micromotors incorporating two enzymes combine weak gradient-dependent chemotactic steering with strong propulsion driven by a spatially uniform fuel. This strategy enables sustained directional transport even when the steering gradient alone would produce only weak migration.

Together, our findings establish a fully aqueous and potentially scalable strategy for generating micromotors with spontaneously emerging, spatially programmed functionalities,  opening avenues for integrated propulsion and directional control.

\section{Materials and Methods}

\subsection*{Simulation methods}

\subsection*{(i) Nanoparticle dynamics in microcondensates: Marangoni-driven transport and
interfacial retention (inside the evaporating droplet)}

We simulate nanoparticle dynamics near the condensate interface using
overdamped Langevin dynamics in the background Marangoni flow $\ub$ and
interfacial trapping force $\Fb_i^{\rm trap}$ [Eq.~\eqref{eq:ftrap}]. Assuming a spherical condensate
of radius $r_c$ centred at the origin, the position $\rb_i$ and magnetic moment
$\pb_i = p \phat_i$ of the $i$th magnetic nanoparticle evolve as

\begin{align}
\frac{d\rb_i}{dt} &=   \ub(\rb_i) + \mu_{\rm{tr}}\left(\Fb_i^{\rm trap}
                     + \Fb_i^{\rm dip} + \Fb_i^{\rm steric}\right)
                     + \sqrt{2D_{\rm{tr}}}\,\xib_i^{\rm tr},
                     \label{eq:drdt_np}\\[10pt]                    
\frac{d\pb_i}{dt} &= \left[\frac{1}{2}\nablab\times\ub(\xb)\big|_{\xb=\rb_i}
                     + \mu_{\rm{rot}}\Tb_i^{\rm dip}
                     + \sqrt{2D_{\rm{rot}}}\,\xib_i^{\rm rot}\right]\times\pb_i,
                     \label{eq:dpdt_np}
\end{align}

where $\mu_{\rm{tr(rot)}}$ is the translational (rotational) mobility of a
nanoparticle of radius $a_{\rm np}$ in a fluid of viscosity $\eta$, and
$D_{\rm{tr(rot)}}=\mu_{\rm{tr(rot)}}k_BT$ is the corresponding diffusion
constant at temperature $T$ \cite{kim1991microhydrodynamics}; $\xib_i^{\rm tr,rot}$  are
unit-variance Gaussian white noises.
%The Marangoni flow $\ub$ and derivation of
%$\Fb_i^{\rm trap}$ are given in Supplementary Information.%
 The net flow field $\ub$ and derivation of
$\Fb_i^{\rm trap}$ are given in Supplementary Information. The dipolar force and torque are
$\Fb_i^{\rm dip}=\sum_{j\ne i}\nablab(\pb_i\cdot\Bb_j)$ and $\Tb_i^{\rm
dip}=\sum_{j\ne i}(\pb_i\times\Bb_j)$, respectively, where $\Bb_j=k_{\rm
mag}(3\phat_j\cdot\rhat_{ij}\rhat_{ij}-\phat_j)/ (3p r_{ij}^3)$ is the field of
dipole $j$ at $\rb_i$ with $k_{\rm mag}=3\mu_0p^2/(4\pi)$, and
$\rb_{ij}=\rb_i-\rb_j$. Here, $\mu_0$ is the magnetic permeability, taken to be
identical inside and outside the condensate for simplicity. Steric interactions
are modelled using the standard Weeks--Chandler--Andersen (WCA) potential
\cite{weeks1971role}. For nonmagnetic nanoparticles, we set $p=0$, rendering the magnetic
interactions in Eqs.~\eqref{eq:drdt_np} \& \eqref{eq:dpdt_np} irrelevant.

In the simulations, we fix $r_c=1\,\mu\mathrm{m}$, $a_{\rm
np}=0.05\,\mu\mathrm{m}$, $\eta=1.235$~mPas, and temperature 25$~^\circ C$, while varying the remaining parameters as specified in the main text and Supplementary Information. We integrate Eqs.~\eqref{eq:drdt_np}--\eqref{eq:dpdt_np} using Euler's method with time step $\Delta t=10^{-6}\,\mathrm{s}$. For simplicity, we consider initial
$\rb_i,\pb_i$ configurations in the condensate's equatorial plane parallel to
the substrate; the flow fields retain their full $3d$ hydrodynamic structure,
while noise-induced out-of-plane fluctuations are negligible at high Péclet
number.

\subsection*{(ii) Particle dynamics: torque-guided active propulsion (in liquid bath)}

The dynamics of the chemically driven, functionalized particles (crosslinked condensates) in a magnetic field $\Bb=B\Bhat$ are described by the overdamped Langevin equations

\begin{align}
\frac{d\rb}{dt} &= v_0\ehat+\tilde{\mu}_{\rm tr}\Fb^{\rm mag}
+\sqrt{2\tilde{D}_{\rm tr}}\,\xib^{\rm tr},
\label{eq:drdt}\\[10pt]
\frac{d\ehat}{dt} &=
\left[\tilde{\mu}_{\rm rot}\Tb^{\rm mag}
+\sqrt{2\tilde{D}_{\rm rot}}\,\xib^{\rm rot}\right]\times\ehat,
\label{eq:dedt}
\end{align}

where $\tilde{\mu}_{\rm tr,rot}$ and $\tilde{D}_{\rm tr,rot}$ are the
translational and rotational mobilities and diffusion constants, respectively,
of crosslinked condensate particles of radius $r_c$, and $\xib^{\rm tr,rot}$ are unit-variance
Gaussian white noises. The magnetic force [Eq.~\eqref{eq_Fm}] is modelled as an effective force 
$\Fb^{\rm mag}=(v_m/\tilde{\mu}_{\rm tr})\Bhat$, where $v_m$ is the velocity with only a magnetic field gradient present, while the magnetic torque comprises polar and apolar aligning
contributions,

$\Tb^{\rm mag}=\alpha_{\rm pol}(\ehat\times\Bhat)
+\alpha_{\rm apol}(\ehat\cdot\Bhat)(\ehat\times\Bhat)$.

Unlike for the nanoparticles, chemical activity drives
$\tilde{D}_{\rm tr,rot}$ far from their thermal values. As shown in the Supplementary Information,
at long times ($\tau\gg\tau_{\rm eff}$) and strong fields, Eqs.~\eqref{eq:drdt}
and~\eqref{eq:dedt} yield the condensate's mean-squared displacement (MSD)

\begin{align}
\left\langle\left|\Delta\rb(t)\right|^2\right\rangle
&\approx 6\tilde{D}_{\rm tr}t
+\left[v_0\left(1-\frac{\tilde{D}_{\rm rot}}
{\tilde{\mu}_{\rm rot}\alpha_{\rm pol}}\right)+v_m\right]^2t^2,
\label{eq:MSD}
\end{align}

where $\tau_{\rm eff}$ is the effective rotational-diffusion timescale. Here we neglected the weak apolar contribution. In the chemical amplification of magnetophoresis, we
extract $\tilde{D}_{\rm tr,rot}$ and $\alpha_{\rm pol}$ by fitting this
expression to the experimental MSD [Fig.~\ref{fig4}~B], using the measured
$v_0$ and $v_m$.  These parameters are then used to integrate
Eqs.~\eqref{eq:drdt} and \eqref{eq:dedt} using Euler's method to obtain the
condensate trajectories shown in Fig.~\ref{fig4}~F; the corresponding
parameters are listed in Table~\ref{tab:sim_params}.

\begin{table}[htbp]
\small
\caption{The fitting in Fig.\ref{fig4}~E is obtained with the following values. }
\label{tbl}
\centering
\begin{tabularx}{\columnwidth}{l l X}
\toprule
\textbf{Parameter} & \textbf{Urea alone} & \textbf{Urea + gradient magnetic field} \\
\midrule
$v_0$      & $6.35~\mu\rm{m}/\rm{s}$        & $6.35~\mu\rm{m}/\rm{s}$\\
$v_m$      & $0~\mu\rm{m}/\rm{s}$          & $0.45~\mu \rm{m}/\rm{s}$\\
$\alpha$   & $0$                           & $3.27$\\
$\tilde{D}_{tr}$      & $0.73~\mu\mathrm{m}^2/\rm{s}$  & $0.92~\mu\rm{m}^2/\rm{s}$\\
$\tilde{D}_{rot}$      & $21.4~\rm{rad}^2/\rm{s}$       & $21.42~\rm{rad}^2/\rm{s}$\\

\midrule
%\bottomrule
\end{tabularx}
\label{tab:sim_params}
\end{table}
%-------------------------

\subsection*{Sample preparation}

Stock solutions of dextran (DEX, M$\mathrm{w}$ = 40000 g/mol) and polyethylene
glycol (PEG, M$\mathrm{w}$ = 6000 g/mol) were prepared at concentrations
ranging from 1–10 wt$\%$ in deionized water. 

To prepare the chitosan solution, glacial acetic acid was first diluted in deionized water at a 1:100 ratio. Chitosan was then dissolved in the diluted acetic acid to obtain solutions of three different concentrations: 0.01, 0.1, and 1 wt$\%$. Each chitosan solution was subsequently added to the PEG–DEX mixture at a final concentration of 1 wt$\%$.

In order to make BSA particles, 5, 100, 300, and 500 $\mu$M of BSA in 20 wt\% PEG, 200 $\mu$M KCl salt and 100 $\mu$M Kp buffer were separately mixed via vortex. (1- 1.5 wt$\%$) Ammonia was added to the above mixture with 500 $\mu$M BSA in the form of ammonium hydroxide to prevent spontaneous phase separation and to achieve tiny microdroplets. 

For functionalization of the particles, we prepared (i) 3 mg/ml of urease solution,  (ii) 0.3 wt $\%$ of catalase solution in phosphate buffer (pH 7), (iii) 2 wt $\%$ solution of {magnetite nanoparticles} (< 0.2 $\mu$m) in deionized water. We mixed 1 wt $\%$ of each solution into the evaporating droplet mixture, as required.  

DNA from E.\textit{coli} (tagged with propidium iodide) (1 mL of 7.38 ng/$\mu$L  DNA with 1 mM propidium iodide) was also added in BSA solution (300 $\mu$M) to be encapsulated in the BSA phases.

\subsection*{Fluorescent tagging}
For fluorescence imaging, polymers and proteins were labeled with fluorescein isothiocyanate (FITC; $\lambda_{em}$ = 517 nm), rhodamine B isothiocyanate (RBITC; $\lambda_{em}$ = 590 nm) or tetramethylrhodamine isothiocyanate (TRITC; $\lambda_{em}$ = 576 nm).

For tagging RBITC with proteins, 0.5 mL of RBITC in dimethyl sulfoxide (DMSO) at 3mg/mL was prepared. The solution was then added dropwise to the protein solution (4 mg/mL in   Na$_{2}$CO$_{3}$- NaHC$ O_{3}$ buffer (pH~9, 50 mM). The reaction mixture was stirred for 4 hours at room temperature. The reaction mixture was then neutralized with NH$_{4}$Cl and dialyzed with a 12~kDa cellulose tubing in PBS buffer (pH~7, 50 mM) to remove unreacted RBITC. 

For tagging with FITC with enzymes, 0.5 mL of FITC (2 mg/mL) in  Na$_{2}$CO$_{3}$- NaHCO$_{3}$ (pH~9, 50 mM) was reacted with 4 mg/mL of enzyme (also in the buffer) for 2 hours in the dark with stirring. And purified by dialysis with a 12 kDa cellulose membrane in PBS buffer (pH ~7, 50 mM).

\subsection*{Enzyme immobilization}
Amine-functionalized polystyrene (PS) particles (1 $\mu$m) were incubated with 3 wt\% glutaraldehyde in PBS for 4 hrs under continuous mixing. The particles were then washed three times with PBS and incubated with 3 mg mL$^{-1}$ FITC-labeled urease for 12 hrs to immobilize the enzyme on the particle surface. Finally, the particles were washed to remove unbound enzyme and stored in PBS at 4 $^\circ$C until use.

\subsection*{Droplet evaporation}
In the experiments, we evaporated droplets of volume 1-10 $\mu$l, deposited on a dry glass slide cleaned with ethanol and water, at laboratory conditions (25 $^\circ$C and 50 $\pm$ 2 \% humidity). Glass slides were treated with oxygen plasma to decrease the contact angle.

\subsection*{Curing the microdroplets}

The DEX + CS microcondensates were collected in a tube containing glutaraldehyde (2 wt\%) in deionized water and kept for 2 hrs.  When using BSA condensates, we added 20 wt\% PEG to the crosslinking solution.  The cured particles were washed with PBS buffer and stored at 4 $^\circ$C. 
During glutaraldehyde treatment, the aldehyde groups react with amine groups to form Schiff base linkages, thereby crosslinking the structures and rendering them mechanically stable and resistant to acidic conditions \cite{beppu2007crosslinking, li2013synthesis}.  The curing process reduces the initial condensate size by approximately 30\% as they form particles.

\subsection*{Alginate particle synthesis} 
The reaction mixture contained 2~wt\% alginate and 20~wt\% PEG. Upon evaporation, alginate condensates with diameters ranging from 1 to 3 $\mu$m were formed. After complete phase separation, 4~wt\% calcium chloride in 20~wt\% PEG solution was added to the evaporated droplets and allowed to crosslink for 20 minutes. The crosslinked condensates were collected and washed with PBS. The functionalization was performed by dispersing the enzymes initially in the drop and allowing to be encapsulated, similar to the case of BSA and DEX-CS.

\subsection*{Imaging}
Both the evaporating droplet, the cross-linked particles, and their active motion were imaged using an inverted fluorescence microscope (Nikon Eclipse Ti2) and cameras (Nikon DS-Qi2 and Vision Research Lab-110) at a maximum resolution of 0.33 $\mu$m/pix with a 90X objective at a frame rate ranging from 50 to 200 fps. Confocal imaging of the cross-linked particles was also performed using an Olympus FV3000 confocal laser scanning microscope. 

\subsection*{Activity measurements with magnetic steering}

Urea and H$_2$O$_2$ solutions were prepared in PBS at concentrations of 0 to 800~mM and 0.1 - 1.5 wt\%, respectively. The functional microparticles were dispersed in the corresponding substrate solutions, and their trajectories were recorded. 

A uniform magnetic field of 10 mT was generated using a Helmholtz coil with a 5 cm radius. The homogeneity of the field was tested using commercially available spherical magnetic beads with a diameter of 5~$\mu$m, which exhibited no noticeable motion.  To flip the particle direction in a uniform field, a permanent magnet was brought briefly within about 1 cm of the solution, producing a field of about 15 mT.

\section*{Conflicts of interest}
IISER Tirupati has filed two patent applications related to the technology described in the manuscript, with the authors (DM, SPP, AS) named as inventors (Indian Patent Application No. 202641049376 and 202541112975). The authors declare no other conflicts of interest. \medskip

\section*{Data availability}
The data supporting this article have been included as part of the Supplementary Information.

\section*{Acknowledgments}
DM and DD acknowledge STARS-2/2023-0752 grant from the Ministry of Education. \medskip

\section*{Author contributions} 
DM conceived the experiments. SPP, AS performed the experiments and analyzed the data, with equal contribution. Ambareesh Shrivastav and TCA performed the theoretical modelling. DD contributed to developing the project.  All authors contributed to writing and reviewing the manuscript.

%%%REFERENCES%%%
\bibliography{References} %You need to replace "rsc" on this line with the name of your .bib file

\providecommand*{\mcitethebibliography}{\thebibliography}
\csname @ifundefined\endcsname{endmcitethebibliography}
{\let\endmcitethebibliography\endthebibliography}{}
\begin{mcitethebibliography}{82}
\providecommand*{\natexlab}[1]{#1}
\providecommand*{\mciteSetBstSublistMode}[1]{}
\providecommand*{\mciteSetBstMaxWidthForm}[2]{}
\providecommand*{\mciteBstWouldAddEndPuncttrue}
  {\def\EndOfBibitem{\unskip.}}
\providecommand*{\mciteBstWouldAddEndPunctfalse}
  {\let\EndOfBibitem\relax}
\providecommand*{\mciteSetBstMidEndSepPunct}[3]{}
\providecommand*{\mciteSetBstSublistLabelBeginEnd}[3]{}
\providecommand*{\EndOfBibitem}{}
\mciteSetBstSublistMode{f}
\mciteSetBstMaxWidthForm{subitem}
{(\emph{\alph{mcitesubitemcount}})}
\mciteSetBstSublistLabelBeginEnd{\mcitemaxwidthsubitemform\space}
{\relax}{\relax}

\bibitem[Note1()]{Note1}
\protect \dag ~Equal contribution\relax
\mciteBstWouldAddEndPuncttrue
\mciteSetBstMidEndSepPunct{\mcitedefaultmidpunct}
{\mcitedefaultendpunct}{\mcitedefaultseppunct}\relax
\EndOfBibitem
\bibitem[Pantarotto \emph{et~al.}(2008)Pantarotto, Browne, and
  Feringa]{pantarotto2008autonomous}
D.~Pantarotto, W.~R. Browne and B.~L. Feringa, \emph{Chemical communications},
  2008,  1533--1535\relax
\mciteBstWouldAddEndPuncttrue
\mciteSetBstMidEndSepPunct{\mcitedefaultmidpunct}
{\mcitedefaultendpunct}{\mcitedefaultseppunct}\relax
\EndOfBibitem
\bibitem[Patra \emph{et~al.}(2013)Patra, Sengupta, Duan, Zhang, Pavlick, and
  Sen]{patra2013intelligent}
D.~Patra, S.~Sengupta, W.~Duan, H.~Zhang, R.~Pavlick and A.~Sen,
  \emph{Nanoscale}, 2013, \textbf{5}, 1273--1283\relax
\mciteBstWouldAddEndPuncttrue
\mciteSetBstMidEndSepPunct{\mcitedefaultmidpunct}
{\mcitedefaultendpunct}{\mcitedefaultseppunct}\relax
\EndOfBibitem
\bibitem[Arqu{\'e} \emph{et~al.}(2022)Arqu{\'e}, Patino, and
  S{\'a}nchez]{arque2022enzyme}
X.~Arqu{\'e}, T.~Patino and S.~S{\'a}nchez, \emph{Chemical Science}, 2022,
  \textbf{13}, 9128--9146\relax
\mciteBstWouldAddEndPuncttrue
\mciteSetBstMidEndSepPunct{\mcitedefaultmidpunct}
{\mcitedefaultendpunct}{\mcitedefaultseppunct}\relax
\EndOfBibitem
\bibitem[Li \emph{et~al.}(2023)Li, Peng, Yan, Mao, Ma, Wilson, He, and
  Tu]{li2023medical}
H.~Li, F.~Peng, X.~Yan, C.~Mao, X.~Ma, D.~A. Wilson, Q.~He and Y.~Tu,
  \emph{Acta Pharmaceutica Sinica B}, 2023, \textbf{13}, 517--541\relax
\mciteBstWouldAddEndPuncttrue
\mciteSetBstMidEndSepPunct{\mcitedefaultmidpunct}
{\mcitedefaultendpunct}{\mcitedefaultseppunct}\relax
\EndOfBibitem
\bibitem[Sim{\'o} \emph{et~al.}(2024)Sim{\'o}, Serra-Casablancas, Hortelao,
  Di~Carlo, Guallar-Garrido, Plaza-Garc{\'\i}a, Rabanal, Ramos-Cabrer,
  Yag{\"u}e, Aguado,\emph{et~al.}]{simo2024urease}
C.~Sim{\'o}, M.~Serra-Casablancas, A.~C. Hortelao, V.~Di~Carlo,
  S.~Guallar-Garrido, S.~Plaza-Garc{\'\i}a, R.~M. Rabanal, P.~Ramos-Cabrer,
  B.~Yag{\"u}e, L.~Aguado \emph{et~al.}, \emph{Nature Nanotechnology}, 2024,
  \textbf{19}, 554--564\relax
\mciteBstWouldAddEndPuncttrue
\mciteSetBstMidEndSepPunct{\mcitedefaultmidpunct}
{\mcitedefaultendpunct}{\mcitedefaultseppunct}\relax
\EndOfBibitem
\bibitem[Chen \emph{et~al.}(2025)Chen, Fan, Fischer, Ghosh, G{\"o}pfrich,
  Golestanian, Hess, Ma, Nelson,
  Pati{\~n}o~Padial,\emph{et~al.}]{chen2025roadmap}
S.~Chen, D.~E. Fan, P.~Fischer, A.~Ghosh, K.~G{\"o}pfrich, R.~Golestanian,
  H.~Hess, X.~Ma, B.~J. Nelson, T.~Pati{\~n}o~Padial \emph{et~al.},
  \emph{Nature nanotechnology}, 2025, \textbf{20}, 990--1000\relax
\mciteBstWouldAddEndPuncttrue
\mciteSetBstMidEndSepPunct{\mcitedefaultmidpunct}
{\mcitedefaultendpunct}{\mcitedefaultseppunct}\relax
\EndOfBibitem
\bibitem[Choi \emph{et~al.}(2024)Choi, Jeong, Sim{\'o}, Bakenecker, Liop, Lee,
  Kim, Kwak, Koh, S{\'a}nchez,\emph{et~al.}]{choi2024urease}
H.~Choi, S.-h. Jeong, C.~Sim{\'o}, A.~Bakenecker, J.~Liop, H.~S. Lee, T.~Y.
  Kim, C.~Kwak, G.~Y. Koh, S.~S{\'a}nchez \emph{et~al.}, \emph{Nature
  communications}, 2024, \textbf{15}, 9934\relax
\mciteBstWouldAddEndPuncttrue
\mciteSetBstMidEndSepPunct{\mcitedefaultmidpunct}
{\mcitedefaultendpunct}{\mcitedefaultseppunct}\relax
\EndOfBibitem
\bibitem[Peters \emph{et~al.}(2014)Peters, Marguet, Marais, Fraaije, van Hest,
  and Lecommandoux]{peters2014cascade}
R.~J. Peters, M.~Marguet, S.~Marais, M.~W. Fraaije, J.~C. van Hest and
  S.~Lecommandoux, \emph{Angewandte chemie}, 2014, \textbf{126}, 150--154\relax
\mciteBstWouldAddEndPuncttrue
\mciteSetBstMidEndSepPunct{\mcitedefaultmidpunct}
{\mcitedefaultendpunct}{\mcitedefaultseppunct}\relax
\EndOfBibitem
\bibitem[Wang \emph{et~al.}(2018)Wang, Zhao, Liu, Shao, Bian, and
  Zhao]{wang2018biomimetic}
H.~Wang, Z.~Zhao, Y.~Liu, C.~Shao, F.~Bian and Y.~Zhao, \emph{Science
  Advances}, 2018, \textbf{4}, eaat2816\relax
\mciteBstWouldAddEndPuncttrue
\mciteSetBstMidEndSepPunct{\mcitedefaultmidpunct}
{\mcitedefaultendpunct}{\mcitedefaultseppunct}\relax
\EndOfBibitem
\bibitem[Wang \emph{et~al.}(2023)Wang, Chen, Jiang, Chen, Xiong, Chen, Xu, Gao,
  Xu, Zhou,\emph{et~al.}]{wang2023cell}
Q.~Wang, K.~Chen, H.~Jiang, C.~Chen, C.~Xiong, M.~Chen, J.~Xu, X.~Gao, S.~Xu,
  H.~Zhou \emph{et~al.}, \emph{Nature Communications}, 2023, \textbf{14},
  5338\relax
\mciteBstWouldAddEndPuncttrue
\mciteSetBstMidEndSepPunct{\mcitedefaultmidpunct}
{\mcitedefaultendpunct}{\mcitedefaultseppunct}\relax
\EndOfBibitem
\bibitem[Demarchi \emph{et~al.}(2023)Demarchi, Goychuk, Maryshev, and
  Frey]{PRL130.128401}
L.~Demarchi, A.~Goychuk, I.~Maryshev and E.~Frey, \emph{Phys. Rev. Lett.},
  2023, \textbf{130}, 128401\relax
\mciteBstWouldAddEndPuncttrue
\mciteSetBstMidEndSepPunct{\mcitedefaultmidpunct}
{\mcitedefaultendpunct}{\mcitedefaultseppunct}\relax
\EndOfBibitem
\bibitem[Zwicker \emph{et~al.}(2017)Zwicker, Seyboldt, Weber, Hyman, and
  Jülicher]{ZwickerNatPhy}
D.~Zwicker, R.~Seyboldt, C.~A. Weber, A.~A. Hyman and F.~Jülicher,
  \emph{Nature Physics}, 2017, \textbf{13}, 408–413\relax
\mciteBstWouldAddEndPuncttrue
\mciteSetBstMidEndSepPunct{\mcitedefaultmidpunct}
{\mcitedefaultendpunct}{\mcitedefaultseppunct}\relax
\EndOfBibitem
\bibitem[Raghavendra \emph{et~al.}(2026)Raghavendra, Shelke, van~der Ham, and
  Vutukuri]{Vutukuri2026program}
T.~S. Raghavendra, Y.~Shelke, S.~van~der Ham and H.~R. Vutukuri, \emph{Commun
  Phys}, 2026, \textbf{9}, 166\relax
\mciteBstWouldAddEndPuncttrue
\mciteSetBstMidEndSepPunct{\mcitedefaultmidpunct}
{\mcitedefaultendpunct}{\mcitedefaultseppunct}\relax
\EndOfBibitem
\bibitem[Cao \emph{et~al.}(2022)Cao, Liu, Ma, Ma, Zu, Sun, Dai, Duan, and
  Xiao]{cao2022oral}
Y.~Cao, S.~Liu, Y.~Ma, L.~Ma, M.~Zu, J.~Sun, F.~Dai, L.~Duan and B.~Xiao,
  \emph{Small}, 2022, \textbf{18}, 2203466\relax
\mciteBstWouldAddEndPuncttrue
\mciteSetBstMidEndSepPunct{\mcitedefaultmidpunct}
{\mcitedefaultendpunct}{\mcitedefaultseppunct}\relax
\EndOfBibitem
\bibitem[Serra-Casablancas \emph{et~al.}(2024)Serra-Casablancas, Di~Carlo,
  Esporrin-Ubieto, Prado-Morales, Bakenecker, and Sanchez]{serra2024catalase}
M.~Serra-Casablancas, V.~Di~Carlo, D.~Esporrin-Ubieto, C.~Prado-Morales, A.~C.
  Bakenecker and S.~Sanchez, \emph{ACS nano}, 2024, \textbf{18},
  16701--16714\relax
\mciteBstWouldAddEndPuncttrue
\mciteSetBstMidEndSepPunct{\mcitedefaultmidpunct}
{\mcitedefaultendpunct}{\mcitedefaultseppunct}\relax
\EndOfBibitem
\bibitem[Dey \emph{et~al.}(2015)Dey, Zhao, Tansi, M{\'e}ndez-Ortiz,
  C{\'o}rdova-Figueroa, Golestanian, and Sen]{dey2015micromotors}
K.~K. Dey, X.~Zhao, B.~M. Tansi, W.~J. M{\'e}ndez-Ortiz, U.~M.
  C{\'o}rdova-Figueroa, R.~Golestanian and A.~Sen, \emph{Nano letters}, 2015,
  \textbf{15}, 8311--8315\relax
\mciteBstWouldAddEndPuncttrue
\mciteSetBstMidEndSepPunct{\mcitedefaultmidpunct}
{\mcitedefaultendpunct}{\mcitedefaultseppunct}\relax
\EndOfBibitem
\bibitem[Arqu{\'e} \emph{et~al.}(2019)Arqu{\'e}, Romero-Rivera, Feixas,
  Pati{\~n}o, Osuna, and S{\'a}nchez]{arque2019intrinsic}
X.~Arqu{\'e}, A.~Romero-Rivera, F.~Feixas, T.~Pati{\~n}o, S.~Osuna and
  S.~S{\'a}nchez, \emph{Nature Communications}, 2019, \textbf{10}, 2826\relax
\mciteBstWouldAddEndPuncttrue
\mciteSetBstMidEndSepPunct{\mcitedefaultmidpunct}
{\mcitedefaultendpunct}{\mcitedefaultseppunct}\relax
\EndOfBibitem
\bibitem[Ghosh \emph{et~al.}(2023)Ghosh, Ghosh, Chatterjee, Bera, Mampallil,
  Ghosh, and Das]{ghosh2023dual}
C.~Ghosh, S.~Ghosh, A.~Chatterjee, P.~Bera, D.~Mampallil, P.~Ghosh and D.~Das,
  \emph{Nature Communications}, 2023, \textbf{14}, 5903\relax
\mciteBstWouldAddEndPuncttrue
\mciteSetBstMidEndSepPunct{\mcitedefaultmidpunct}
{\mcitedefaultendpunct}{\mcitedefaultseppunct}\relax
\EndOfBibitem
\bibitem[Tang \emph{et~al.}(2026)Tang, Han, Ma, Patel, Gong, Zhang,
  Criado-Hidalgo, Yoo, Li, Kim,\emph{et~al.}]{tang2026enzymatic}
S.~Tang, H.~Han, X.~Ma, P.~N. Patel, C.~Gong, J.~Zhang, E.~Criado-Hidalgo,
  J.~Yoo, J.~Li, G.~Kim \emph{et~al.}, \emph{Nature Nanotechnology}, 2026,
  1--10\relax
\mciteBstWouldAddEndPuncttrue
\mciteSetBstMidEndSepPunct{\mcitedefaultmidpunct}
{\mcitedefaultendpunct}{\mcitedefaultseppunct}\relax
\EndOfBibitem
\bibitem[Zhang \emph{et~al.}(2021)Zhang, Li, Gao, Fan, Pang, Li, Wu, Xie, and
  He]{zhang2021dual}
H.~Zhang, Z.~Li, C.~Gao, X.~Fan, Y.~Pang, T.~Li, Z.~Wu, H.~Xie and Q.~He,
  \emph{Science Robotics}, 2021, \textbf{6}, eaaz9519\relax
\mciteBstWouldAddEndPuncttrue
\mciteSetBstMidEndSepPunct{\mcitedefaultmidpunct}
{\mcitedefaultendpunct}{\mcitedefaultseppunct}\relax
\EndOfBibitem
\bibitem[Naikwadi \emph{et~al.}(2024)Naikwadi, Paul, Biswas, Chitlange, and
  Wavhale]{naikwadi2024self}
N.~Naikwadi, M.~Paul, S.~Biswas, S.~Chitlange and R.~Wavhale, \emph{Scientific
  Reports}, 2024, \textbf{14}, 31547\relax
\mciteBstWouldAddEndPuncttrue
\mciteSetBstMidEndSepPunct{\mcitedefaultmidpunct}
{\mcitedefaultendpunct}{\mcitedefaultseppunct}\relax
\EndOfBibitem
\bibitem[Kirillova \emph{et~al.}(2019)Kirillova, Marschelke, and
  Synytska]{kirillova2019hybrid}
A.~Kirillova, C.~Marschelke and A.~Synytska, \emph{ACS applied materials \&
  interfaces}, 2019, \textbf{11}, 9643--9671\relax
\mciteBstWouldAddEndPuncttrue
\mciteSetBstMidEndSepPunct{\mcitedefaultmidpunct}
{\mcitedefaultendpunct}{\mcitedefaultseppunct}\relax
\EndOfBibitem
\bibitem[Joye and McClements(2014)]{joye2014biopolymer}
I.~J. Joye and D.~J. McClements, \emph{Current opinion in colloid \& interface
  science}, 2014, \textbf{19}, 417--427\relax
\mciteBstWouldAddEndPuncttrue
\mciteSetBstMidEndSepPunct{\mcitedefaultmidpunct}
{\mcitedefaultendpunct}{\mcitedefaultseppunct}\relax
\EndOfBibitem
\bibitem[Carvalho \emph{et~al.}(2023)Carvalho, Dos~Santos, Polli~Silvestre,
  Tavares, Chorilli, and Daflon~Gremi{\~a}o]{carvalho2023multifunctional}
S.~G. Carvalho, A.~M. Dos~Santos, A.~L. Polli~Silvestre, A.~G. Tavares,
  M.~Chorilli and M.~P. Daflon~Gremi{\~a}o, \emph{Expert Opinion on Drug
  Delivery}, 2023, \textbf{20}, 1231--1249\relax
\mciteBstWouldAddEndPuncttrue
\mciteSetBstMidEndSepPunct{\mcitedefaultmidpunct}
{\mcitedefaultendpunct}{\mcitedefaultseppunct}\relax
\EndOfBibitem
\bibitem[Zhou \emph{et~al.}(2021)Zhou, Mayorga-Martinez, Pan{\'e}, Zhang, and
  Pumera]{zhou2021magnetically}
H.~Zhou, C.~C. Mayorga-Martinez, S.~Pan{\'e}, L.~Zhang and M.~Pumera,
  \emph{Chemical Reviews}, 2021, \textbf{121}, 4999--5041\relax
\mciteBstWouldAddEndPuncttrue
\mciteSetBstMidEndSepPunct{\mcitedefaultmidpunct}
{\mcitedefaultendpunct}{\mcitedefaultseppunct}\relax
\EndOfBibitem
\bibitem[Naz \emph{et~al.}(2024)Naz, Zhang, Chen, Yang, Dou, Mann, and
  Li]{naz2024self}
M.~Naz, L.~Zhang, C.~Chen, S.~Yang, H.~Dou, S.~Mann and J.~Li,
  \emph{Communications chemistry}, 2024, \textbf{7}, 79\relax
\mciteBstWouldAddEndPuncttrue
\mciteSetBstMidEndSepPunct{\mcitedefaultmidpunct}
{\mcitedefaultendpunct}{\mcitedefaultseppunct}\relax
\EndOfBibitem
\bibitem[Park \emph{et~al.}(2025)Park, Ding, and Schuster]{park2025pompoms}
A.~S. Park, E.~A. Ding and B.~S. Schuster, \emph{Biomacromolecules}, 2025\relax
\mciteBstWouldAddEndPuncttrue
\mciteSetBstMidEndSepPunct{\mcitedefaultmidpunct}
{\mcitedefaultendpunct}{\mcitedefaultseppunct}\relax
\EndOfBibitem
\bibitem[Wakata \emph{et~al.}(2025)Wakata, Li, and Sun]{wakata2025evaporation}
Y.~Wakata, M.~Li and C.~Sun, \emph{Current Opinion in Colloid \& Interface
  Science}, 2025,  101987\relax
\mciteBstWouldAddEndPuncttrue
\mciteSetBstMidEndSepPunct{\mcitedefaultmidpunct}
{\mcitedefaultendpunct}{\mcitedefaultseppunct}\relax
\EndOfBibitem
\bibitem[Mampallil(2025)]{mampallil2025evaporation}
D.~Mampallil, \emph{Current Opinion in Colloid \& Interface Science}, 2025,
  \textbf{78}, 101938\relax
\mciteBstWouldAddEndPuncttrue
\mciteSetBstMidEndSepPunct{\mcitedefaultmidpunct}
{\mcitedefaultendpunct}{\mcitedefaultseppunct}\relax
\EndOfBibitem
\bibitem[Guo \emph{et~al.}(2021)Guo, Kinghorn, Zhang, Li, Poonam, Tanner, and
  Shum]{Guo2021}
W.~Guo, A.~B. Kinghorn, Y.~Zhang, Q.~Li, A.~D. Poonam, J.~A. Tanner and H.~C.
  Shum, \emph{Nature communications}, 2021, \textbf{12}, 1--13\relax
\mciteBstWouldAddEndPuncttrue
\mciteSetBstMidEndSepPunct{\mcitedefaultmidpunct}
{\mcitedefaultendpunct}{\mcitedefaultseppunct}\relax
\EndOfBibitem
\bibitem[May \emph{et~al.}(2022)May, Hartmann, and Hardt]{May2022}
A.~May, J.~Hartmann and S.~Hardt, \emph{Soft Matter}, 2022, \textbf{18},
  6313--6317\relax
\mciteBstWouldAddEndPuncttrue
\mciteSetBstMidEndSepPunct{\mcitedefaultmidpunct}
{\mcitedefaultendpunct}{\mcitedefaultseppunct}\relax
\EndOfBibitem
\bibitem[Watanabe and Yanagisawa(2022)]{watanabe2022evaporation}
C.~Watanabe and M.~Yanagisawa, \emph{Life}, 2022, \textbf{12}, 373\relax
\mciteBstWouldAddEndPuncttrue
\mciteSetBstMidEndSepPunct{\mcitedefaultmidpunct}
{\mcitedefaultendpunct}{\mcitedefaultseppunct}\relax
\EndOfBibitem
\bibitem[Rai \emph{et~al.}(2024)Rai, Gopu, Parameswaran, Adhyapak, and
  Mampallil]{Rai2024}
R.~Rai, M.~Gopu, S.~P. Parameswaran, T.~C. Adhyapak and D.~Mampallil,
  \emph{Soft Matter}, 2024, \textbf{20}, 8260--8266\relax
\mciteBstWouldAddEndPuncttrue
\mciteSetBstMidEndSepPunct{\mcitedefaultmidpunct}
{\mcitedefaultendpunct}{\mcitedefaultseppunct}\relax
\EndOfBibitem
\bibitem[Tan \emph{et~al.}(2016)Tan, Diddens, Lv, Kuerten, Zhang, and
  Lohse]{pnas.1602260113}
H.~Tan, C.~Diddens, P.~Lv, J.~G.~M. Kuerten, X.~Zhang and D.~Lohse,
  \emph{Proceedings of the National Academy of Sciences}, 2016, \textbf{113},
  8642--8647\relax
\mciteBstWouldAddEndPuncttrue
\mciteSetBstMidEndSepPunct{\mcitedefaultmidpunct}
{\mcitedefaultendpunct}{\mcitedefaultseppunct}\relax
\EndOfBibitem
\bibitem[Kumar \emph{et~al.}(2024)Kumar, Gopu, Parameswaran, Joshi, and
  Mampallil]{kumar2024evaporative}
M.~Kumar, M.~Gopu, S.~P. Parameswaran, P.~Joshi and D.~Mampallil, \emph{JCIS
  Open}, 2024, \textbf{13}, 100101\relax
\mciteBstWouldAddEndPuncttrue
\mciteSetBstMidEndSepPunct{\mcitedefaultmidpunct}
{\mcitedefaultendpunct}{\mcitedefaultseppunct}\relax
\EndOfBibitem
\bibitem[Ivanov \emph{et~al.}(2025)Ivanov, Cao, Schaaf, Caire~da Silva, and
  Landfester]{ivanov2025dynamic}
T.~Ivanov, S.~Cao, M.~Schaaf, L.~Caire~da Silva and K.~Landfester,
  \emph{Chemical Communications}, 2025, \textbf{61}, 15227--15230\relax
\mciteBstWouldAddEndPuncttrue
\mciteSetBstMidEndSepPunct{\mcitedefaultmidpunct}
{\mcitedefaultendpunct}{\mcitedefaultseppunct}\relax
\EndOfBibitem
\bibitem[Dave \emph{et~al.}(2025)Dave, Kassem, Coste, Xu, Tayarani-Najjaran,
  Podbev{\v{s}}ek, Colon-De~Leon, Zhang, Ortuno~Macias,
  Sementa,\emph{et~al.}]{dave2025adaptive}
D.~R. Dave, S.~Kassem, M.~Coste, L.~Xu, M.~Tayarani-Najjaran,
  D.~Podbev{\v{s}}ek, P.~Colon-De~Leon, S.~Zhang, L.~Ortuno~Macias, D.~Sementa
  \emph{et~al.}, \emph{Nature Materials}, 2025, \textbf{24}, 1465--1475\relax
\mciteBstWouldAddEndPuncttrue
\mciteSetBstMidEndSepPunct{\mcitedefaultmidpunct}
{\mcitedefaultendpunct}{\mcitedefaultseppunct}\relax
\EndOfBibitem
\bibitem[Keating(2012)]{keating2012aqueous}
C.~D. Keating, \emph{Accounts of chemical research}, 2012, \textbf{45},
  2114--2124\relax
\mciteBstWouldAddEndPuncttrue
\mciteSetBstMidEndSepPunct{\mcitedefaultmidpunct}
{\mcitedefaultendpunct}{\mcitedefaultseppunct}\relax
\EndOfBibitem
\bibitem[Ramos \emph{et~al.}(2022)Ramos, Bernard, Ganachaud, and
  Miserez]{ramos2022protein}
R.~Ramos, J.~Bernard, F.~Ganachaud and A.~Miserez, \emph{Small Science}, 2022,
  \textbf{2}, 2100095\relax
\mciteBstWouldAddEndPuncttrue
\mciteSetBstMidEndSepPunct{\mcitedefaultmidpunct}
{\mcitedefaultendpunct}{\mcitedefaultseppunct}\relax
\EndOfBibitem
\bibitem[Sakuta \emph{et~al.}(2020)Sakuta, Fujita, Hamada, Hayashi, Takiguchi,
  Tsumoto, and Yoshikawa]{sakuta2020self}
H.~Sakuta, F.~Fujita, T.~Hamada, M.~Hayashi, K.~Takiguchi, K.~Tsumoto and
  K.~Yoshikawa, \emph{ChemBioChem}, 2020, \textbf{21}, 3323--3328\relax
\mciteBstWouldAddEndPuncttrue
\mciteSetBstMidEndSepPunct{\mcitedefaultmidpunct}
{\mcitedefaultendpunct}{\mcitedefaultseppunct}\relax
\EndOfBibitem
\bibitem[Qi \emph{et~al.}(2024)Qi, Ma, Zeng, Huang, Zhang, Deng, Kong, and
  Liu]{Qi2024}
C.~Qi, X.~Ma, Q.~Zeng, Z.~Huang, S.~Zhang, X.~Deng, T.~Kong and Z.~Liu,
  \emph{Nat Commun}, 2024, \textbf{15}, 1107\relax
\mciteBstWouldAddEndPuncttrue
\mciteSetBstMidEndSepPunct{\mcitedefaultmidpunct}
{\mcitedefaultendpunct}{\mcitedefaultseppunct}\relax
\EndOfBibitem
\bibitem[Hu \emph{et~al.}(2023)Hu, Cheng, Chen, Xin, and Wu]{hu2023liquid}
Z.~Hu, L.~Cheng, Q.~Chen, T.~Xin and X.~Wu, \emph{Materials Advances}, 2023,
  \textbf{4}, 5643--5652\relax
\mciteBstWouldAddEndPuncttrue
\mciteSetBstMidEndSepPunct{\mcitedefaultmidpunct}
{\mcitedefaultendpunct}{\mcitedefaultseppunct}\relax
\EndOfBibitem
\bibitem[Vashist \emph{et~al.}(2026)Vashist, Pal, and
  Ghosh]{vashist2026magnetically}
E.~Vashist, B.~Pal and A.~Ghosh, \emph{Advanced Functional Materials}, 2026,
  \textbf{36}, e27479\relax
\mciteBstWouldAddEndPuncttrue
\mciteSetBstMidEndSepPunct{\mcitedefaultmidpunct}
{\mcitedefaultendpunct}{\mcitedefaultseppunct}\relax
\EndOfBibitem
\bibitem[Sengupta \emph{et~al.}(2013)Sengupta, Dey, Muddana, Tabouillot, Ibele,
  Butler, and Sen]{sengupta2013enzyme}
S.~Sengupta, K.~K. Dey, H.~S. Muddana, T.~Tabouillot, M.~E. Ibele, P.~J. Butler
  and A.~Sen, \emph{Journal of the American Chemical Society}, 2013,
  \textbf{135}, 1406--1414\relax
\mciteBstWouldAddEndPuncttrue
\mciteSetBstMidEndSepPunct{\mcitedefaultmidpunct}
{\mcitedefaultendpunct}{\mcitedefaultseppunct}\relax
\EndOfBibitem
\bibitem[Meng \emph{et~al.}(2025)Meng, Tang, and Li]{meng2025micro}
X.~Meng, Y.~Tang and Q.~Li, \emph{Responsive Materials}, 2025, \textbf{3},
  e20250021\relax
\mciteBstWouldAddEndPuncttrue
\mciteSetBstMidEndSepPunct{\mcitedefaultmidpunct}
{\mcitedefaultendpunct}{\mcitedefaultseppunct}\relax
\EndOfBibitem
\bibitem[Chen \emph{et~al.}(2024)Chen, Prado-Morales,
  S{\'a}nchez-deAlc{\'a}zar, and S{\'a}nchez]{chen2024enzymatic}
S.~Chen, C.~Prado-Morales, D.~S{\'a}nchez-deAlc{\'a}zar and S.~S{\'a}nchez,
  \emph{Journal of Materials Chemistry B}, 2024, \textbf{12}, 2711--2719\relax
\mciteBstWouldAddEndPuncttrue
\mciteSetBstMidEndSepPunct{\mcitedefaultmidpunct}
{\mcitedefaultendpunct}{\mcitedefaultseppunct}\relax
\EndOfBibitem
\bibitem[Wu \emph{et~al.}(2022)Wu, Folio, Zhu, Jang, Chen, Feng, Gambardella,
  Sort, Puigmart{\'\i}-Luis, Ergeneman,\emph{et~al.}]{wu2022motion}
J.~Wu, D.~Folio, J.~Zhu, B.~Jang, X.~Chen, J.~Feng, P.~Gambardella, J.~Sort,
  J.~Puigmart{\'\i}-Luis, O.~Ergeneman \emph{et~al.}, \emph{Advanced
  Intelligent Systems}, 2022, \textbf{4}, 2200192\relax
\mciteBstWouldAddEndPuncttrue
\mciteSetBstMidEndSepPunct{\mcitedefaultmidpunct}
{\mcitedefaultendpunct}{\mcitedefaultseppunct}\relax
\EndOfBibitem
\bibitem[Villa \emph{et~al.}(2018)Villa, Krej{\v{c}}ov{\'a}, Novotn{\`y},
  Heger, Sofer, and Pumera]{villa2018cooperative}
K.~Villa, L.~Krej{\v{c}}ov{\'a}, F.~Novotn{\`y}, Z.~Heger, Z.~Sofer and
  M.~Pumera, \emph{Advanced Functional Materials}, 2018, \textbf{28},
  1804343\relax
\mciteBstWouldAddEndPuncttrue
\mciteSetBstMidEndSepPunct{\mcitedefaultmidpunct}
{\mcitedefaultendpunct}{\mcitedefaultseppunct}\relax
\EndOfBibitem
\bibitem[Unruh \emph{et~al.}(2023)Unruh, Savage, and Sen]{unruh2023remote}
A.~Unruh, E.~J. Savage and A.~Sen, \emph{Chemistry of Materials}, 2023,
  \textbf{35}, 10099--10105\relax
\mciteBstWouldAddEndPuncttrue
\mciteSetBstMidEndSepPunct{\mcitedefaultmidpunct}
{\mcitedefaultendpunct}{\mcitedefaultseppunct}\relax
\EndOfBibitem
\bibitem[Lu \emph{et~al.}(2016)Lu, Liu, Oh, Gargava, Kendall, Nie, DeVoe, and
  Raghavan]{lu2016catalytic}
A.~X. Lu, Y.~Liu, H.~Oh, A.~Gargava, E.~Kendall, Z.~Nie, D.~L. DeVoe and S.~R.
  Raghavan, \emph{ACS applied materials \& interfaces}, 2016, \textbf{8},
  15676--15683\relax
\mciteBstWouldAddEndPuncttrue
\mciteSetBstMidEndSepPunct{\mcitedefaultmidpunct}
{\mcitedefaultendpunct}{\mcitedefaultseppunct}\relax
\EndOfBibitem
\bibitem[Jambon-Puillet \emph{et~al.}(2024)Jambon-Puillet, Testa, Lorenz,
  Style, Rebane, and Dufresne]{jambon2024phase}
E.~Jambon-Puillet, A.~Testa, C.~Lorenz, R.~W. Style, A.~A. Rebane and E.~R.
  Dufresne, \emph{Nature communications}, 2024, \textbf{15}, 3919\relax
\mciteBstWouldAddEndPuncttrue
\mciteSetBstMidEndSepPunct{\mcitedefaultmidpunct}
{\mcitedefaultendpunct}{\mcitedefaultseppunct}\relax
\EndOfBibitem
\bibitem[Patel \emph{et~al.}(2022)Patel, Singh, Saini, and
  Mukherjee]{patel2022macromolecular}
C.~K. Patel, S.~Singh, B.~Saini and T.~K. Mukherjee, \emph{The Journal of
  Physical Chemistry Letters}, 2022, \textbf{13}, 3636--3644\relax
\mciteBstWouldAddEndPuncttrue
\mciteSetBstMidEndSepPunct{\mcitedefaultmidpunct}
{\mcitedefaultendpunct}{\mcitedefaultseppunct}\relax
\EndOfBibitem
\bibitem[Zhang \emph{et~al.}(2021)Zhang, Contini, Hindley, Bolognesi, Elani,
  and Ces]{zhang2021engineering}
S.~Zhang, C.~Contini, J.~W. Hindley, G.~Bolognesi, Y.~Elani and O.~Ces,
  \emph{Nature communications}, 2021, \textbf{12}, 1673\relax
\mciteBstWouldAddEndPuncttrue
\mciteSetBstMidEndSepPunct{\mcitedefaultmidpunct}
{\mcitedefaultendpunct}{\mcitedefaultseppunct}\relax
\EndOfBibitem
\bibitem[Barbosa \emph{et~al.}(2013)Barbosa, Ortiz, Berenguer-Murcia, Torres,
  Rodrigues, and Fernandez-Lafuente]{barbosa2013glutaraldehyde}
O.~Barbosa, C.~Ortiz, A.~Berenguer-Murcia, R.~Torres, R.~C. Rodrigues and
  R.~Fernandez-Lafuente, \emph{Rsc Advances}, 2013, \textbf{4},
  1583--1600\relax
\mciteBstWouldAddEndPuncttrue
\mciteSetBstMidEndSepPunct{\mcitedefaultmidpunct}
{\mcitedefaultendpunct}{\mcitedefaultseppunct}\relax
\EndOfBibitem
\bibitem[Ma \emph{et~al.}(2016)Ma, Sun, Hargrove, Chen, Song, Dong, Lu, Fan,
  Fu, and Lei]{ma2016biocompatible}
X.~Ma, X.~Sun, D.~Hargrove, J.~Chen, D.~Song, Q.~Dong, X.~Lu, T.-H. Fan, Y.~Fu
  and Y.~Lei, \emph{Scientific reports}, 2016, \textbf{6}, 19370\relax
\mciteBstWouldAddEndPuncttrue
\mciteSetBstMidEndSepPunct{\mcitedefaultmidpunct}
{\mcitedefaultendpunct}{\mcitedefaultseppunct}\relax
\EndOfBibitem
\bibitem[Wang \emph{et~al.}(2013)Wang, Duan, Ahmed, Mallouk, and
  Sen]{wang2013small}
W.~Wang, W.~Duan, S.~Ahmed, T.~E. Mallouk and A.~Sen, \emph{Nano Today}, 2013,
  \textbf{8}, 531--554\relax
\mciteBstWouldAddEndPuncttrue
\mciteSetBstMidEndSepPunct{\mcitedefaultmidpunct}
{\mcitedefaultendpunct}{\mcitedefaultseppunct}\relax
\EndOfBibitem
\bibitem[Zhao \emph{et~al.}(2018)Zhao, Gentile, Mohajerani, and
  Sen]{zhao2018powering}
X.~Zhao, K.~Gentile, F.~Mohajerani and A.~Sen, \emph{Accounts of chemical
  research}, 2018, \textbf{51}, 2373--2381\relax
\mciteBstWouldAddEndPuncttrue
\mciteSetBstMidEndSepPunct{\mcitedefaultmidpunct}
{\mcitedefaultendpunct}{\mcitedefaultseppunct}\relax
\EndOfBibitem
\bibitem[Pati{\~n}o \emph{et~al.}(2018)Pati{\~n}o, Feiner-Gracia, Arqu{\'e},
  Miguel-L{\'o}pez, Jannasch, Stumpp, Sch{\"a}ffer, Albertazzi, and
  S{\'a}nchez]{patino2018influence}
T.~Pati{\~n}o, N.~M. Feiner-Gracia, X.~Arqu{\'e}, A.~Miguel-L{\'o}pez,
  A.~Jannasch, T.~Stumpp, E.~Sch{\"a}ffer, L.~Albertazzi and S.~S{\'a}nchez,
  \emph{Journal of the American Chemical Society}, 2018, \textbf{140},
  7896--7903\relax
\mciteBstWouldAddEndPuncttrue
\mciteSetBstMidEndSepPunct{\mcitedefaultmidpunct}
{\mcitedefaultendpunct}{\mcitedefaultseppunct}\relax
\EndOfBibitem
\bibitem[Golestanian \emph{et~al.}(2005)Golestanian, Liverpool, and
  Ajdari]{golestanian2005propulsion}
R.~Golestanian, T.~B. Liverpool and A.~Ajdari, \emph{Physical review letters},
  2005, \textbf{94}, 220801\relax
\mciteBstWouldAddEndPuncttrue
\mciteSetBstMidEndSepPunct{\mcitedefaultmidpunct}
{\mcitedefaultendpunct}{\mcitedefaultseppunct}\relax
\EndOfBibitem
\bibitem[Chen \emph{et~al.}(2013)Chen, Zhang, Dang, and Shu]{chen2013The}
H.~Chen, Q.~Zhang, Y.~Dang and G.~Shu, \emph{Advance Journal of Food Science
  and Technology}, 2013, \textbf{5}, 932--935\relax
\mciteBstWouldAddEndPuncttrue
\mciteSetBstMidEndSepPunct{\mcitedefaultmidpunct}
{\mcitedefaultendpunct}{\mcitedefaultseppunct}\relax
\EndOfBibitem
\bibitem[Liang \emph{et~al.}(2005)Liang, Feng, Meng, and Liang]{LiangUrease}
Z.~Liang, Y.~Feng, S.~Meng and Z.~Liang, \emph{Chinese Chemical Letters}, 2005,
  \textbf{16}, 135--138\relax
\mciteBstWouldAddEndPuncttrue
\mciteSetBstMidEndSepPunct{\mcitedefaultmidpunct}
{\mcitedefaultendpunct}{\mcitedefaultseppunct}\relax
\EndOfBibitem
\bibitem[Giraldo-Zuniga \emph{et~al.}(2006)Giraldo-Zuniga, dos Reis~Coimbra,
  Arquete, Minim, Mendes~da Silva, and Maffia]{Cristinasurfacetension2006}
A.~D. Giraldo-Zuniga, J.~S. dos Reis~Coimbra, D.~A. Arquete, L.~A. Minim, L.~H.
  Mendes~da Silva and M.~C. Maffia, \emph{Journal of Chemical \& Engineering
  Data}, 2006, \textbf{51}, 1144--1147\relax
\mciteBstWouldAddEndPuncttrue
\mciteSetBstMidEndSepPunct{\mcitedefaultmidpunct}
{\mcitedefaultendpunct}{\mcitedefaultseppunct}\relax
\EndOfBibitem
\bibitem[Binks and Horozov(2006)]{Binks_Horozov_2006}
B.~P. Binks and T.~S. Horozov, \emph{Colloidal Particles at Liquid Interfaces},
  Cambridge University Press, 2006, pp. 1--10\relax
\mciteBstWouldAddEndPuncttrue
\mciteSetBstMidEndSepPunct{\mcitedefaultmidpunct}
{\mcitedefaultendpunct}{\mcitedefaultseppunct}\relax
\EndOfBibitem
\bibitem[Svelto(2010)]{laserssvelto2010}
O.~Svelto, \emph{Principles of Lasers}, Springer US, 2010\relax
\mciteBstWouldAddEndPuncttrue
\mciteSetBstMidEndSepPunct{\mcitedefaultmidpunct}
{\mcitedefaultendpunct}{\mcitedefaultseppunct}\relax
\EndOfBibitem
\bibitem[Luo \emph{et~al.}(2020)Luo, Li, Wan, Yang, Chen, and
  Guan]{luo2020enhanced}
M.~Luo, S.~Li, J.~Wan, C.~Yang, B.~Chen and J.~Guan, \emph{Langmuir}, 2020,
  \textbf{36}, 7005--7013\relax
\mciteBstWouldAddEndPuncttrue
\mciteSetBstMidEndSepPunct{\mcitedefaultmidpunct}
{\mcitedefaultendpunct}{\mcitedefaultseppunct}\relax
\EndOfBibitem
\bibitem[Suwa \emph{et~al.}(2023)Suwa, Tsukahara, and
  Watarai]{suwa2023applications}
M.~Suwa, S.~Tsukahara and H.~Watarai, \emph{Lab on a Chip}, 2023, \textbf{23},
  1097--1127\relax
\mciteBstWouldAddEndPuncttrue
\mciteSetBstMidEndSepPunct{\mcitedefaultmidpunct}
{\mcitedefaultendpunct}{\mcitedefaultseppunct}\relax
\EndOfBibitem
\bibitem[Bailey \emph{et~al.}(2022)Bailey, Sprenger, Grillo, L\"owen, and
  Isa]{PREL052602}
M.~R. Bailey, A.~R. Sprenger, F.~Grillo, H.~L\"owen and L.~Isa, \emph{Phys.
  Rev. E}, 2022, \textbf{106}, L052602\relax
\mciteBstWouldAddEndPuncttrue
\mciteSetBstMidEndSepPunct{\mcitedefaultmidpunct}
{\mcitedefaultendpunct}{\mcitedefaultseppunct}\relax
\EndOfBibitem
\bibitem[Popescu \emph{et~al.}(2018)Popescu, Uspal, Bechinger, and
  Fischer]{PopescuNanolett}
M.~N. Popescu, W.~E. Uspal, C.~Bechinger and P.~Fischer, \emph{Nano Letters},
  2018, \textbf{18}, 5345--5349\relax
\mciteBstWouldAddEndPuncttrue
\mciteSetBstMidEndSepPunct{\mcitedefaultmidpunct}
{\mcitedefaultendpunct}{\mcitedefaultseppunct}\relax
\EndOfBibitem
\bibitem[Vinze \emph{et~al.}(2021)Vinze, Choudhary, and
  Pushpavanam]{vinze2021motion}
P.~M. Vinze, A.~Choudhary and S.~Pushpavanam, \emph{Physics of Fluids}, 2021,
  \textbf{33}, 032011\relax
\mciteBstWouldAddEndPuncttrue
\mciteSetBstMidEndSepPunct{\mcitedefaultmidpunct}
{\mcitedefaultendpunct}{\mcitedefaultseppunct}\relax
\EndOfBibitem
\bibitem[Lee \emph{et~al.}(2026)Lee, Choo, Chaudhary, Park, Chang, Byun, Kim,
  and Lee]{lee2026benchmarking}
H.~S. Lee, S.~Choo, M.~Chaudhary, J.~H. Park, J.~Chang, Y.~Byun, K.-T. Kim and
  J.-Y. Lee, \emph{Small Methods}, 2026, \textbf{10}, e01671\relax
\mciteBstWouldAddEndPuncttrue
\mciteSetBstMidEndSepPunct{\mcitedefaultmidpunct}
{\mcitedefaultendpunct}{\mcitedefaultseppunct}\relax
\EndOfBibitem
\bibitem[Roberts \emph{et~al.}(2020)Roberts, Miao, Costa, Simon, Kelly, Shah,
  Zauscher, and Chilkoti]{roberts2020complex}
S.~Roberts, V.~Miao, S.~Costa, J.~Simon, G.~Kelly, T.~Shah, S.~Zauscher and
  A.~Chilkoti, \emph{Nature communications}, 2020, \textbf{11}, 1342\relax
\mciteBstWouldAddEndPuncttrue
\mciteSetBstMidEndSepPunct{\mcitedefaultmidpunct}
{\mcitedefaultendpunct}{\mcitedefaultseppunct}\relax
\EndOfBibitem
\bibitem[Wang \emph{et~al.}(2021)Wang, Sun, Xu, Du, Zhu, and
  Wu]{wang2021mechanism}
Y.~Wang, R.~Sun, X.~Xu, M.~Du, B.~Zhu and C.~Wu, \emph{Food \& function}, 2021,
  \textbf{12}, 12696--12705\relax
\mciteBstWouldAddEndPuncttrue
\mciteSetBstMidEndSepPunct{\mcitedefaultmidpunct}
{\mcitedefaultendpunct}{\mcitedefaultseppunct}\relax
\EndOfBibitem
\bibitem[Chen \emph{et~al.}(2023)Chen, Yu, Zhang, Wu, Ma, and Li]{chen2023co}
X.~Chen, C.~Yu, Y.~Zhang, Y.-C. Wu, Y.~Ma and H.-J. Li, \emph{Food \&
  Function}, 2023, \textbf{14}, 3169--3178\relax
\mciteBstWouldAddEndPuncttrue
\mciteSetBstMidEndSepPunct{\mcitedefaultmidpunct}
{\mcitedefaultendpunct}{\mcitedefaultseppunct}\relax
\EndOfBibitem
\bibitem[Yu \emph{et~al.}(2017)Yu, Nguyen, Cheow, and Hadinoto]{yu2017new}
H.~Yu, M.-H. Nguyen, W.~S. Cheow and K.~Hadinoto, \emph{Materials Science and
  Engineering: C}, 2017, \textbf{75}, 25--33\relax
\mciteBstWouldAddEndPuncttrue
\mciteSetBstMidEndSepPunct{\mcitedefaultmidpunct}
{\mcitedefaultendpunct}{\mcitedefaultseppunct}\relax
\EndOfBibitem
\bibitem[Zhang \emph{et~al.}(2025)Zhang, Wang, Lu, Quan, Wang, Song, and
  Guo]{zhang2025advanced}
Y.~Zhang, Y.~Wang, Y.~Lu, H.~Quan, Y.~Wang, S.~Song and H.~Guo, \emph{Journal
  of Nanobiotechnology}, 2025, \textbf{23}, 400\relax
\mciteBstWouldAddEndPuncttrue
\mciteSetBstMidEndSepPunct{\mcitedefaultmidpunct}
{\mcitedefaultendpunct}{\mcitedefaultseppunct}\relax
\EndOfBibitem
\bibitem[Abbaspourrad \emph{et~al.}(2013)Abbaspourrad, Datta, and
  Weitz]{abbaspourrad2013controlling}
A.~Abbaspourrad, S.~S. Datta and D.~A. Weitz, \emph{Langmuir}, 2013,
  \textbf{29}, 12697--12702\relax
\mciteBstWouldAddEndPuncttrue
\mciteSetBstMidEndSepPunct{\mcitedefaultmidpunct}
{\mcitedefaultendpunct}{\mcitedefaultseppunct}\relax
\EndOfBibitem
\bibitem[Zheng \emph{et~al.}(2025)Zheng, Guo, Ma, Li, and
  Shao]{zheng2025recent}
M.~Zheng, M.~Guo, F.~Ma, W.~Li and Y.~Shao, \emph{Nanoscale Advances}, 2025,
  \textbf{7}, 4780--4802\relax
\mciteBstWouldAddEndPuncttrue
\mciteSetBstMidEndSepPunct{\mcitedefaultmidpunct}
{\mcitedefaultendpunct}{\mcitedefaultseppunct}\relax
\EndOfBibitem
\bibitem[Kim and Karrila(1991)]{kim1991microhydrodynamics}
S.~Kim and S.~J. Karrila, \emph{Microhydrodynamics: Principles and Selected
  Applications}, Butterworth-Heinemann, Boston, 1991\relax
\mciteBstWouldAddEndPuncttrue
\mciteSetBstMidEndSepPunct{\mcitedefaultmidpunct}
{\mcitedefaultendpunct}{\mcitedefaultseppunct}\relax
\EndOfBibitem
\bibitem[Weeks \emph{et~al.}(1971)Weeks, Chandler, and Andersen]{weeks1971role}
J.~D. Weeks, D.~Chandler and H.~C. Andersen, \emph{The Journal of chemical
  physics}, 1971, \textbf{54}, 5237--5247\relax
\mciteBstWouldAddEndPuncttrue
\mciteSetBstMidEndSepPunct{\mcitedefaultmidpunct}
{\mcitedefaultendpunct}{\mcitedefaultseppunct}\relax
\EndOfBibitem
\bibitem[Beppu \emph{et~al.}(2007)Beppu, Vieira, Aimoli, and
  Santana]{beppu2007crosslinking}
M.~Beppu, R.~Vieira, C.~Aimoli and C.~Santana, \emph{Journal of membrane
  science}, 2007, \textbf{301}, 126--130\relax
\mciteBstWouldAddEndPuncttrue
\mciteSetBstMidEndSepPunct{\mcitedefaultmidpunct}
{\mcitedefaultendpunct}{\mcitedefaultseppunct}\relax
\EndOfBibitem
\bibitem[Li \emph{et~al.}(2013)Li, Shan, Zhou, Fang, Wang, Xu, Han, Ibrahim,
  Guo, Xie,\emph{et~al.}]{li2013synthesis}
B.~Li, C.-L. Shan, Q.~Zhou, Y.~Fang, Y.-L. Wang, F.~Xu, L.-R. Han, M.~Ibrahim,
  L.-B. Guo, G.-L. Xie \emph{et~al.}, \emph{Marine drugs}, 2013, \textbf{11},
  1534--1552\relax
\mciteBstWouldAddEndPuncttrue
\mciteSetBstMidEndSepPunct{\mcitedefaultmidpunct}
{\mcitedefaultendpunct}{\mcitedefaultseppunct}\relax
\EndOfBibitem
\end{mcitethebibliography}
\bibliographystyle{rsc} %the RSC's .bst file 
\end{document}